\documentclass[preprint,authoryear,12pt]{elsarticle}

\usepackage{graphicx}
 \usepackage{subfig}

\usepackage{amssymb}
\usepackage{multirow}

\journal{Advances in Space Research}

\begin{document}

\begin{frontmatter}



\title{Time Series Analysis of Active Galactic Nuclei: The case of Arp 102B, 3C 390.3, NGC 5548 and NGC 4051}


\author[label 1] {A.  Kova{\v c}evi{\'c}\corref{cor}}
\address[label 1]{Deprtment of Astronomy, Faculty of Mathematics, University
  of Belgrade, Studentski trg 16, 11000 Belgrade, Serbia}
\cortext[cor]{Corresponding author}

\ead{andjelka@matf.bg.ac.rs}


\author[label 2]{L. \v C. Popovi{\'c}}
\address[label 2]{Astronomical Observatory, Volgina 7, 11060
Belgrade 38, Serbia}
\ead{lpopovic@aob.rs}
\author[label 3]{A. I Shapovalova}
\address[label 3]{Special Astrophysical Observatory of the Russian AS, Nizhnij Arkhyz, 
Karachaevo - Cherkesia 369167, Russia}

\author[label 1]{D. Ili{\' c}}
\ead{dilic@matf.bg.ac.rs}

\author[label 3]{A. N. Burenkov}

\author[label 4]{V. H. Chavushyan}
\address[label 4]{Instituto Nacional de Astrof\'{i}sica, \'{O}ptica y 
Electr\'{o}nica, Apartado Postal 51y 216, 72000 Puebla,  M\'{e}xico}
\begin{abstract}

{
We used the Z-transformed Discrete Correlation Function (ZDCF) and 
the Stochastic Process Estimation for AGN Reverberation (SPEAR) methods for the time series analysis of
the continuum and the H${\alpha}$ and H${\beta}$ line fluxes of a sample of well known type 1 active galactic nuclei 
(AGNs):  Arp 102B, 3C 390.3, NGC 5548, and NGC 4051, where the first two objects are showing
double-peaked emission line profiles. The aim of this work is to compare the time lag measurements 
from these two methods, and check if there is a connection with other emission line properties. 
We found that the obtained time lags from H$\beta$ are larger than those derived from the H$\alpha$ 
analysis for Arp 102B, 3C 390.3 and NGC 5548. This may indicate that the H$\beta$ line 
originates at larger radii in these objects. Moreover, we found that the ZDCF and SPEAR time lags 
are highly correlated ($r \sim0.87$), and that the error ranges of both ZDCF and SPEAR 
time lags are correlated with the FWHM of used emission lines ($r\sim 0.7$). This increases
the uncertainty of the black hole mass estimates using the virial theorem for AGNs with broader lines. 
}

\end{abstract}


\end{frontmatter}

\parindent=0.5 cm

\section{Introduction}

In active galactic nuclei (AGNs),  the variation of the optical broad emission lines and
the continuum flux are correlated, but with a certain time-delay, $\tau$,
 that corresponds to the time propagation across the broad line region 
(BLR).

Therefore, a widely  accepted method for the BLR size determination
is to determine the first moment of transfer 
function (the time-delay or ‘lag’) between the broad line and continuum
light curves using the cross-correlation function (CCF) technique \citep[][]{GS86,GP87}. Measuring time lags 
is important for understanding the physical size of the BLRs, and that
in combination with the virial theorem yields to determination of the mass of the supermassive black 
hole (SMBH) in the center of AGNs.
  
The CCF is a convolution of the delay map (the strength of reprocessed light as a function of 
various time delays) with the continuum auto-correlation function (ACF) \citep{H99}.
With this, delay map could be obtained from the CCF with application of  some deconvolution technique on ACF.
Thus, the lag extracted by cross-correlation analysis depends not only on  the delay distribution 
(e.g. transfer function), but also on the characteristics ACF of the continuum variations.
Because of this, different monitoring programs may provide different time lags even when 
the underlying delay map is the same \citep{H04}. For example, the continuum variability properties of 
AGN NGC 5548 vary from year to year leading to a change in the ACF, and as an artifact of this, 
 a change in the lag could be measured without  a visible change in the delay-map \citep{CH06}.
There are attempts of using  more refined velocity-delay mapping which  aims to recover
the delay map  rather than just a characteristic time lag 
\citep{VHW94,KD95,PW94,HWP91,K03,CH06,DP06,BD06,BW08,DP06}. These mapping methods generally 
require more complete data than the cross-correlation analyses. Nevertheless, there 
are many systematic errors that can affect time lag determinations \citep[see][]{D09, D11, Bentz10},
e.g. errors of the emission line width measurements due to narrow-line contamination, low
data quality (i.e., S/N), or blending of spectral features, undersampling of data 
(simple shortage of data and offset in timescales sampled), etc. Understanding and mitigating 
these systematic uncertainties are important, since they affect the SMBH mass estimates,
 in the AGN reverberation monitoring method (that gives the BLR size and thus the SMBH mass using the
virial theorem) or in a widely used single-epoch BH mass measurement, that is based on the relationship 
between the AGN luminosity and the size of the BLR (for a review on SMBH mass estimates see e.g. \cite{P11}, 
 and reference therein).

There are several commonly accepted  CCF methods for estimations of the time lag between
two data series: the Discrete Correlation Function (DCF) method
\citep{EK88,P93,WP94},  the Interpolated Cross Correlation Function (ICCF), \citet{GP87,P93}, 
 the Modified Cross Correlation Function (MCCF), \citet{KG89,KOS06}, etc. In the theory,
the CCF   requires that time series are uniformly sampled, which is not commonly achievable 
from the ground based observations.

 
 The  discrete sampling problem has been addressed differently in different CCF methods. The ICCF of \citet{GP87} uses 
a linear interpolation scheme to determine the missing data in the light curves, while
the discrete correlation function \citep[DCF;][]{EK88} can utilize a binning scheme to 
approximate the missing data. On the other hand, z-transformed Discrete Correlation Function formulation 
\citep[ZDCF;][]{A97,A13} is also a binning type of method and is a modification of the DCF technique, but 
its distinguishing feature is that the data are binned by equal population rather than equal binwidth as in the DCF. 
Up to now,  there are several studies which showed that the ZDCF is more robust than both ICCF and DCF when 
applied to sparsely and unequally sampled light curves \citep[see e.g.][]{GM99,Roy00}. 
For example, \citet{Liu08,Liu11}  calculated and analized ZDCFs between unequally sampled light curves of AGNs, 
and obtained successfully interband time lags. Consequently, the ZDCF is applicable 
and reliable for the analysis  of the unequally sampled light curves. Therefore,  we will be 
considering this technique in our analysis.

More recently, \citet{Zu11} gave one new method to estimate the time lag between continuum 
and broad emission line of AGNs and this is so-called Stochastic Process Estimation for AGN Reverberation 
or SPEAR method. It uses the assumption that all emission-line light curves are time-delayed, scaled, 
smoothed, and displaced versions of the continuum. This approach fits the light curves directly using a damped 
random walk model (DRW model, \citet{K09,Koz10,ML10} and aligns them to recover the time lag and its 
statistical confidence limits.  \citet{Zu11}  re-measured the time lags in a sample of reverberation-mapped AGNs 
with the SPEAR method and demonstrated its ability to recover accurate time lags. The method has since
been successfully used to improve the reverberation-mapped measurements \citep{Gri12,DP12} and even recover
velocity-delay maps \citep{Gri13}.  For example, \citet{Zh13} applied both SPEAR and CCF methods to
calculate time lags for AGN 3C 390.3, showing that these values are strongly correlated (see their Fig 3).
Thus, the SPEAR method is used as a second one in our analysis of time lags.


 The aim of this paper is to compare the results of the two (ZDCF and SPEAR) methods
that are applied on the four well known type 1 AGNs (Arp 102B, 3C 390.3, NGC 5548, and 
NGC 4051), and discuss the reliability of their time lag determination. Moreover, we will check if there is 
a connection of time lag measurements with other emission line properties, especially the 
emission line width.

Since the diversity in the shape and width of the line profiles implies the diversity in the shape of transfer
function and thus significant variations in the strength of the line response to continuum variations \citep[see][]{R95}, we 
selected our sample to have two types of objects: AGNs with double-peaked broad line profiles (Arp 102B and 3C 390.3) 
and typical type 1 AGNs with single-peaked profiles (NGC 5548 and NGC 4051).  The double-peaked broad lines 
of Arp 102B and 3C390.3 were successfully modeled with the circular relativistic Keplerian 
accretion disk around a SMBH \citep{C89,CH89,Halp90,E97,E09}. However, their broad emission line
profiles vary in a complex way, thus more complex BLR models are needed \citep{Gez07, Pop11}. 
The other two objects, NGC 5548 and NGC 4051 are typical broad single-peaked line
AGNs. {\bf  Over the 6-year monitoring campaign since 1996, NGC 5548 sometimes crosses the boundary between a Sy1 type  and a  Sy1.8  type 
AGN \citep{Sh04}}, and NGC 4051 is relatively nearby object that was part of the extensive reverberation 
mapping campaign \citep[see][]{P04,D06}.  {\bf All the sample AGNs have redshift $z < 0.06$, thus the time dilation correction is negligible.}


The paper is organized as follows: in the section Data  used data samples are described, in 
the section Methodology we describe the ZDCF and SPEAR  methods in more details, in section Results we give
estimates of time lags for all objects,  in section Discussion we discuss our results, and finally in the 
last section we outline our conclusions.

\section{Data}
 
In our analysis we use the integral fluxes of the broad H${\alpha}$ and H${\beta}$ emission 
lines and the fluxes of the red and blue continuum.
Spectra of Arp 102B, 3C 390.3, and NGC 5548 were taken from our monitoring campaign 
\citep[see papers][]{Sh04, Sh10, Sh13} and were observed with the following telescopes:
the 6 m and 1 m telescopes of the Special Astrophysical Observatory (SAO) of the Russian Academy of Science 
(Russia), the INAOE’s 2.1 m telescope of the Guillermo Haro Observatory at Cananea, Sonora, Mexico,
  the 2.1 m telescope of the Observatorio Astronomico Nacional at San Pedro Martir, 
  Baja California, Mexico (only Arp 102B), and the 3.5 m and 2.2 m telescopes of Calar Alto observatory, 
  Spain (only Arp 102B is observed). Taking into account all observations, the mean sampling rate is 
  0.016  observations per day for Arp 102B, for  3C 390.3 it is 0.011  observations per day, while for
  NGC 5548 the mean sampling rate is 0.032 observations per day. 
  The basic information about the sources of spectroscopic observations of these 3 objects are listed in 
  Table \ref{table1}, while more details about the observations and data reduction can be found in \citet{Sh04, Sh10, Sh13}.

\begin{table}
\caption{Sources of spectroscopic observations for Arp 102B, 3C 390.3, and NGC 5548. 
Columns left to the right: Object, Name of the observatory, Telescope aperture and spectrograph, 
Projected spectrograph entrance apertures (slit width$\times$slit length in $^{\prime\prime} \times ^{\prime \prime}$), 
Focus of the telescope, Number or spectra obtained, Observation period (in years).  All of the observations are obtained
 with good transparency.}
\resizebox{14.2cm}{!}{
\begin{tabular}{lllllll}

\hline
Object & Observatory& Tel.aperture&Aperture&Focus&No&Period \\
          &  &+equipment&  $^{\prime\prime} \times ^{\prime \prime}$          &    &   &     \\
\hline
\multirow{16}{*}{Arp 102B}& SAO (Russia)&                 6 m + Long slit&  2.0$\times$6.0& Nasmith&4&  1998-2004 \\
&             & & & & & \\
& SAO (Russia)&         6 m + UAGS&   2.0$\times$6.0&   Prime&11&  1998-2004 \\
&              & & & & & \\
&                SAO (Russia)&         6 m + Scorpio&   1.0$\times$6.07&   Prime&4&  2004-2009 \\
 &                            & & & & & \\
& SAO (Russia)&     1 m + GAD&   4.0$\times$9.45&  Casscgrain&19&   2004-2005 \\
&  & & &&& 2006-2010\\
& Gullermo Haro (Mexico)&          2.1 m + B,C&   2.5$\times$6.0&  Cassegrain&104&   2000-2007\\
&              & & & & & \\ 
&  San Pedro Martir (Mexico)&    2.1 m + B,C&   2.5$\times$6.0&  Cassegrain&9&   2005-2007 \\
&                 & & & & & \\  
&  Calar Alto (Spain)&3.5m+B,C&(1.5-2.1)$\times$3.5 &Cassegrain& 8& 1987-1993 \\
&               &TWIN & & & & \\  
&  Calar Alto (Spain)&2.2m+B,C&2.0$\times$3.5 &Cassegrain&2&1992-1994 \\ 
&                & & & & & \\
\hline
\multirow{8}{*}{3C 390.3}& SAO (Russia)&   6 m + Long slit& 1.0$\times6.1,2.0\times$6.0 & Prime&69&  1995-2007\\
&  & & & & & \\
&  SAO (Russia)&   6 m + Long slith&   2.0$\times$6.0&   Nasmith&13&  1995-1999 \\
&  & & & & & \\ 
& SAO (Russia)&     1 m + GAD&  4.2$\times$19.8, 4.2x13.8&  Cassegrain&2&   1996-1998 \\
&  & & & & & \\
& Gullermo Haro (Mexico)&          2.1 m + B,C &   2.5$\times$6.0&  Cassegrain&73&   1998-2007\\
&  & & & & & \\
\hline
\multirow{6}{*}{NGC 5548}& SAO (Russia)&                 1 m + UAGS& $4.2\times19.8$& Cassagrain&58&  1996-2003\\ 
&  & & & & & \\ 
& SAO (Russia)&         6 m + UAGS&   2.0$\times$6.0&   Nasmith&35&  1996-2001\\
&  & & & & & \\
&Gullermo Haro (Mexico)&          2.1 m + B,C&   2.5$\times$6.0&  Cassegrain&23&   1998-2003\\ 
& & & & & & \\
\hline
\end{tabular}
}
\label{table1}
\end{table}

\begin{figure}
\begin{center}
\includegraphics[width=9cm]{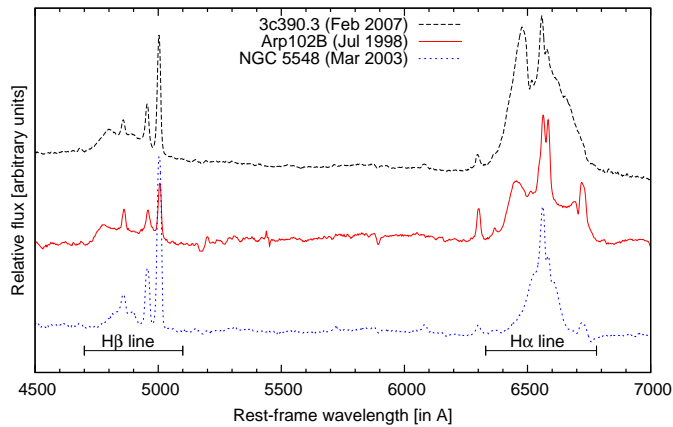}
\caption{Examples of spectra of objects from our monitoring campaign: 3C 390.3, Arp
102B, and NGC 5548. The dates of observations are given in brackets.}
\label{figure1}
\end{center}
\end{figure}

 We show the representative optical spectra of Arp 102B, 3C 390.3, and NGC 5548  
from our monitoring campaining (Fig. \ref{figure1}).  
 In the case of 3C 390.3, the typical wavelength interval covered  was from 4000 \AA \,   to 7500 \AA \,,  
 the spectral resolution varied between 5 and 15 \AA \,, and the $S/N$ ratio was $>$50 in 
 the continuum near H${\alpha}$  and H${\beta}$  lines. The  mean uncertainties in the fluxes are:
 for the continuum $\sim3\%$, for the broad H${\beta}$ line $\sim 5\% $, and for the broad  H${\alpha}$  line 
 $\sim 10\%$.  These quantities were estimated by comparing results from spectra obtained within a time interval shorter than 3 days. 
 For Arp 102B, the typical observed wavelength range was from 4000  \AA \,  to 7500 \AA \,,
 the spectral resolution was in the range of 8–-15  \AA \,, and the $S/N$ ratio was  20--50. 
 For this object we also used the observations taken with the Calar Alto 
 3.5 m and 2.2 m telescopes, for which wavelength ranges from 3630   \AA \, to 9100  \AA \,, and the spectral
resolution was $10–-15$  \AA \,. The mean errors in the observed continuum fluxes at 5100  \AA \, and  6100  \AA \, 
are 3.4$\%$ and 4.4$\%$, respectively, while in the observed fluxes of emission lines H${\alpha}$ 
and H${\beta}$ are $\sim 4\%$ and $\sim 3\%$, respectively. For NGC 5548, the typical wavelength range was
  from 4000  \AA \, to 7500  \AA \,, the spectral resolution was 4.5--15  \AA \,, and the S/N ratio was $>50$.
The mean error in our flux determination for both, the H$\beta$ and the continuum, is 
$\sim3\%$, while it is $\sim5\%$ for H$\alpha$. For these 3 objects, all details of line and continuum 
fluxes variability (e.g. light curves) are given in \citet{Sh04, Sh10, Sh13}.

Finally, for NGC 4051 the data were taken from \citet{D06}. All details of data and 
  discussion of observations and data reduction could be found in \citet{D06}  
  and \citet{P04}.

\section{Methodology}

Before describing in details the two used method, we outline the basics of the CCF analysis, 
that is a standard tool in the time series analysis. 
The CCF is used as a measure of the similarity or correlation between two time series
 (i.e., light curves) as a function of the time shift between them. Its 
 formal description is

\begin{equation}
 \label{eq:1}
   {%
       CCF(\tau)=\frac{ E \left[ (s(t)- \bar{s})(p(t+{\tau})- \bar{p}) \right]}{\sqrt{ 
    V_{s}V_{p}}}
    }
   \end{equation}

where $s(t)$ and $p(t)$ are two light curves, $\tau$ is the time lag,  $\bar {s}$, $\bar{p}$ are 
the means of the corresponding light curves, and ${V}_{s},{V}_{p}$ are the variances of the corresponding 
light curves. The time lag $\tau$ corresponding to the peak in the CCF is quantifying the delay between two time series.

 Eq. (1) assumes an uniform sampling in the light curves, but since this is not usually the case
some approximative methods have to be employed for providing the uniform sampling. A well known approximation 
technique is the ICCF method given by \citet{GP87} \citep[which was later modified 
by][]{WP94,PWB98, P04} based on a linear interpolation scheme which allows to determine the 
missing data in the light curve. In this method the cross-correlation is performed  twice: first  time series
is interpolated and correlated with the second one, and vice versa. The average is taken for the final result. 
So, the form of the ICCF, when the 
interpolation is performed in time series $p(t)$ is given as

\begin{equation}
\label{eq:2}
 {%
    ICCF(\tau)=\frac{1}{N}\sum \frac{ (s_{i}(t)-\bar{s})\left[  L_{p}\left[p(t+{\tau})\right]-\bar{p}\right]}{\sqrt{ 
    {\sigma}_{s}{\sigma}_{p}}}
    }
  \end{equation}  
where $\rm L_p$ is a piecewise linear interpolation of the time series $p$ at $t+\tau$, $N$ is the number of data 
in time series $s$,  and $\sigma_{s}$, $\sigma_{p}$ are standard deviations of the corresponding time series.
 While  $\bar {s}$, $ \bar{p}$ have been already defined 
above.
 The ICCF works well if assuming that the variations in the light curve are smooth.

 Another method, is the DCF \citep{EK88}  which utilizes 
a binning scheme to approximate the missing data.  In this procedure, pairwise combinations of 
measured flux values from the two light curves are binned according to their relative delays or lags.
For the data pairs within each lag bin, a quantity which is essentially a Pearsons linear correlation coefficient 
is computed \citep{PT92}. This technique estimates CCF with 

\begin{equation}
\label{eq:3}
 {%
    DCF(\tau)=\frac{1}{M_{\tau}}\sum_{(i,j)\in K}\frac{ (s_i-\bar s)(p_j-\bar p)}{\sqrt{ \sigma_s \sigma_p}} 
    }
  \end{equation}  

{\bf where $s_{i}$ and $p_{j}$ are pair of i-th and j-th data points from the first and the second light curves,
respectively, in the collection K of i and j within selected bin  $M_{\tau}$ at $\tau-{\frac{\Delta{\tau}}{2}}\leq{t_{j}-t_{i}}\leq \tau+{\frac{\Delta{\tau}}{2}}$,
 and  $\Delta{\tau}$ is a chosen bin size.}
 The errors in the DCF are estimated from the standard deviation in each bin.

Both the ICCF and DCF methods are widely applied in the time series analyses. The algorithms and limitations of the methods have been
 discussed in detail by \citet{RP90} and by \citet{WP94}, and here we just outline that the DCF does not give reasonable results
  for the limited number of data points. Moreover, both methods are unable to provide direct estimates of the uncertainties in 
  the estimated lags,  thus the assumption-dependent Monte Carlo simulations are used.  
Also, it is not clear how such operations as interpolation and rebinning will change the result of the analysis. 

 Further, we describe in details the two CCF techniques applied in this paper.

\subsection{Z-transformed Discrete Correlation Function (ZDCF)}

 Another   common method  used to estimate the CCF of non uniformly sampled light curves is 
 the ZDCF \citep[see][]{A97,A13}. This is also a binning algorithm and can be considered as a modification of 
 the DCF technique, in which all points from the two light curves are ordered according to 
 their time difference $\tau_{ij}$, and binned according to the user’s perception. 
 The ZDCF scheme is based on the approximation  of the cross-correlation function CCF($\tau$) 
 with the correlation coefficient between the two time series. It is defined as: 
  
  \begin{equation}
    \label{eq:4}
 {%
    r=\sum_{i=1}^{n}\frac{ (s_{i}-\bar{s})(p_{i}-\bar{p})}{ {(n+1)S}_{s}{S}_{p}}
    }
  \end{equation}  
where $n$ is the number of time series pairs in a given time lag bin and, in difference to the DCF
 where the normalization is by the mean and standard deviation of the whole time series.
{\bf The variances $S_{s}^{2}$ and $S_{p}^{2}$ } are used to normalize each individual bin. 
Therefore, the ZDCF is normalized by the mean and standard deviation of the cross 
correlated light curves using only the data points that contribute to the 
calculation of each lag, i.e. called 'local CCF' \citep{W99}.
 The method uses the Fisher’s z-transform of $r$ \citep{FF20} to estimate the confidence level 
of a measured correlation, where the binning is defined by the equal population rather than 
the equal width $\Delta{\tau}$. The z-transform’s convergence requires a minimum of $n_{min} = 11$ points per bin \citep{A97,A13}.

For equal binning, the ZDCF estimates of the CCF are equal to the DCF results, but the uncertainties are better behaved.
However, it has been shown that the calculation of the ZDCF is more robust than the ICCF and  DCF methods when applied to 
very sparsely and irregularly sampled light curves  \citep[see e. g.,][]{EA96,AT97,Chi00,GM99,Roy00,Liu08, Liu11,A13}. 
This   method has advantage on our light curves which have more sparse data by nature. 
More details will be discussed in Section 4.1. 
{\bf  The ZDCF method determines the peak of ZDCF profile and the corresponding lag 
uncertainty with maximum likelihood function on a large number of Monte Carlo simulations. }

\subsection{Stochastic Process Estimation for AGN Reverberation (SPEAR)}

Beside CCF methods which measure the time delay between the continuum and emission-line variations,  
 \citet{Zu11}  recently provided an alternative method of measuring 
reverberation time lags called Stochastic Process Estimation for AGN Reverberation 
(SPEAR).

The SPEAR method treats gaps in the temporal coverage of light curves in a well defined  statistical approach.
For any given damped random walk (DRW) model parameters, the stochastic process model not only interpolates 
between data points, but also self-consistently estimates and includes the uncertainties in the interpolation. 
The method can: i) separate light curve means, trends, and systematic errors  from variability signals
 and measurement noise in a self-consistent way, ii) derive lags of multiple emission lines and their covariances simultaneously,
  and iii) provide statistical confidence limits on the lag estimates as well as other parameters. 
 
The fundamental idea of the DRW model is that the variability of signal $s(t)$ (e.g. continuum)
has one simple exponential covariance between two different epochs $t_i$ and $t_j$, in a form
  \begin{equation}
    \label{eq:5}
 {%
     <s(t_i)s(t_j)>={\sigma}^{2} \cdot exp^{\frac{-\left|{t_{i}}-{t_{j}} \right|}{\tau_{0}}}. 
    }
  \end{equation}
Then, through the two DRW parameters, the damped intrinsic variability time-scale $\tau_0$ and the 
damped intrinsic variability amplitude $\sigma$, the AGN variabilities in both observed and unobserved 
epochs can be well reproduced.

The covariance between line  signal $s_l$, and continuum signal $s_c$, can be defined as
\begin{equation}
\label{eq:6}
 {%
     <s_{l}(t_i)s_{c}(t_j)>=  \int dt' g(t_{i}-t')<s_{c}(t')s_{c}(t_j)>  
    }
  \end{equation}  
 Here we assumed that  a simple top hat 
function $g(t-t^{\prime})=A\cdot\Delta \tau, (t-t^{\prime})\in [t_{1},t_{2}]$, which has a mean lag  
$<\tau>=(t_{1}+t_{2})$ and temporal width $\Delta \tau=t_{2}-t_{1}$. 
The scaling coefficient $A$ determines the line response for a given change in the
continuum, but SPEAR largely views it as a nuisance variable.
 
The SPEAR method provides confidence limits for all parameters through calculation of
the Highest Posterior Density (HPD) intervals.  The logarithmic  likelihood value 
of these parameters are calculated.  The ratio log-likelihood functions of this method is defined in the following form 
\begin{equation}
    \label{eq:7}
 {%
     L=-2 \cdot ln (\frac{\mathcal{L}}{\mathcal{L}_{max}})  
    }
  \end{equation}  
The likelihood $\mathcal{L}$ is defined in Eq. (17) in \citet{Zu11} and is proportional to $e^\frac{-\chi^{2}}{2}$. 
The best model, corresponding to $\mathcal{L}_{max}$, is associated with 
the minimum $\chi^2$  of the lag model, thus minimizing L in Eq. 5, as it is shown in \citet{Gri12},  the following formula is obtained 

 \begin{equation}
    \label{eq:8}
 {%
     \Delta \chi ^{2}=- 2 \cdot ln(\frac{\mathcal{L}}{\mathcal{L}_{max}}).  
    }
  \end{equation}   
Based on this considerations, this measures  $\Delta \chi ^{2}$ between models using each lag and the best model.

The SPEAR method has been successfully used to improve reverberation mapping measurements \citep{Gri12, DP12}, providing velocity-delay maps \citep{Gri13}, and the super massive black hole mass estimates  \citep{Gri13b}. 
The difference between the SPEAR and classical  cross-correlation methods are in two basic aspects: 
i) SPEAR explicitly models the light curve and transfer function and fits it to  the continuum and the line data, 
maximizing $\mathcal{L}$ of the model and then computing uncertainties
using the  Markov chain Monte Carlo method; ii) SPEAR models the continuum light curve as an autoregressive process using a
{\bf  DRW model  where  a self-correcting term} is added to a 
 random walk model to confine any deviations back toward the mean value. The parameters of the DRW model are included in the fits and their uncertainties, as it is a 
top-hat model of the transfer function and the light curve means.   
This  auto-regressive process  has been demonstrated using large samples of quasar 
light curves to be a valid statistical representation of quasar variability  \citep{GP87,K09,Koz10, ML10,MHH11,BJ12,AKB13, Zu13}.

\section{Results}
 
\subsection{The ZDCF analysis}

\begin{figure}
\begin{center}
\includegraphics[width = 6.cm]{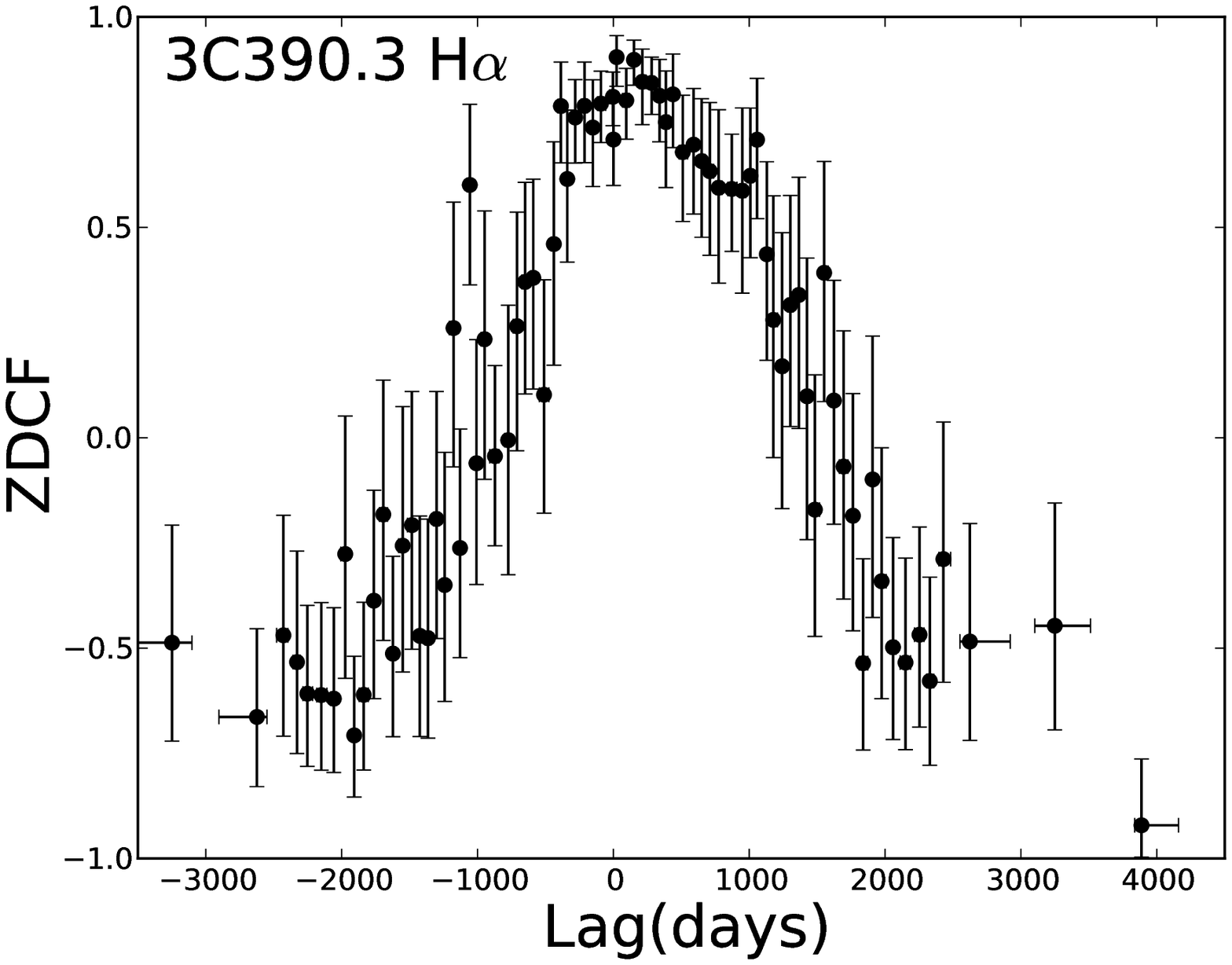}
\includegraphics[width = 6.cm]{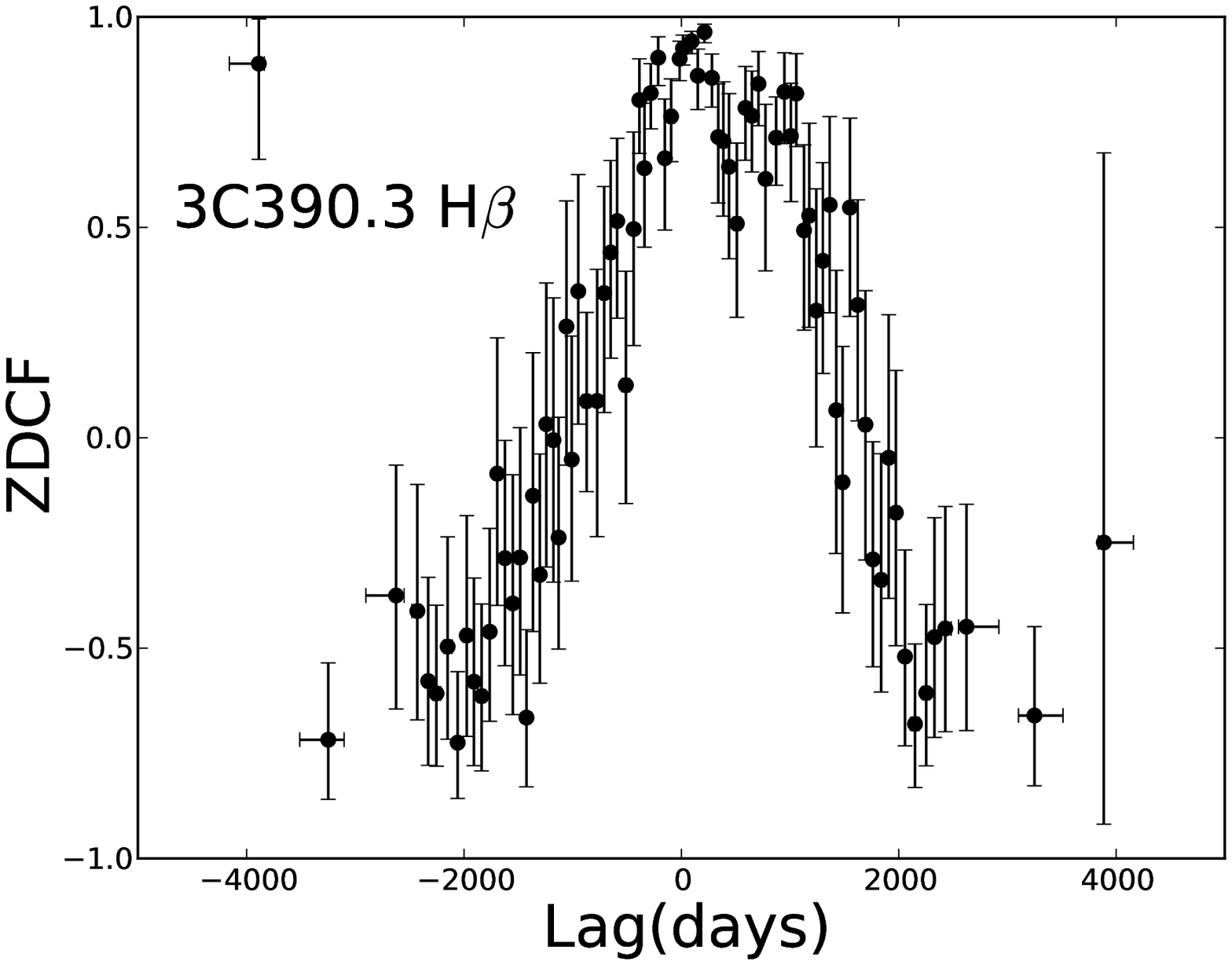}
\includegraphics[width = 6.cm]{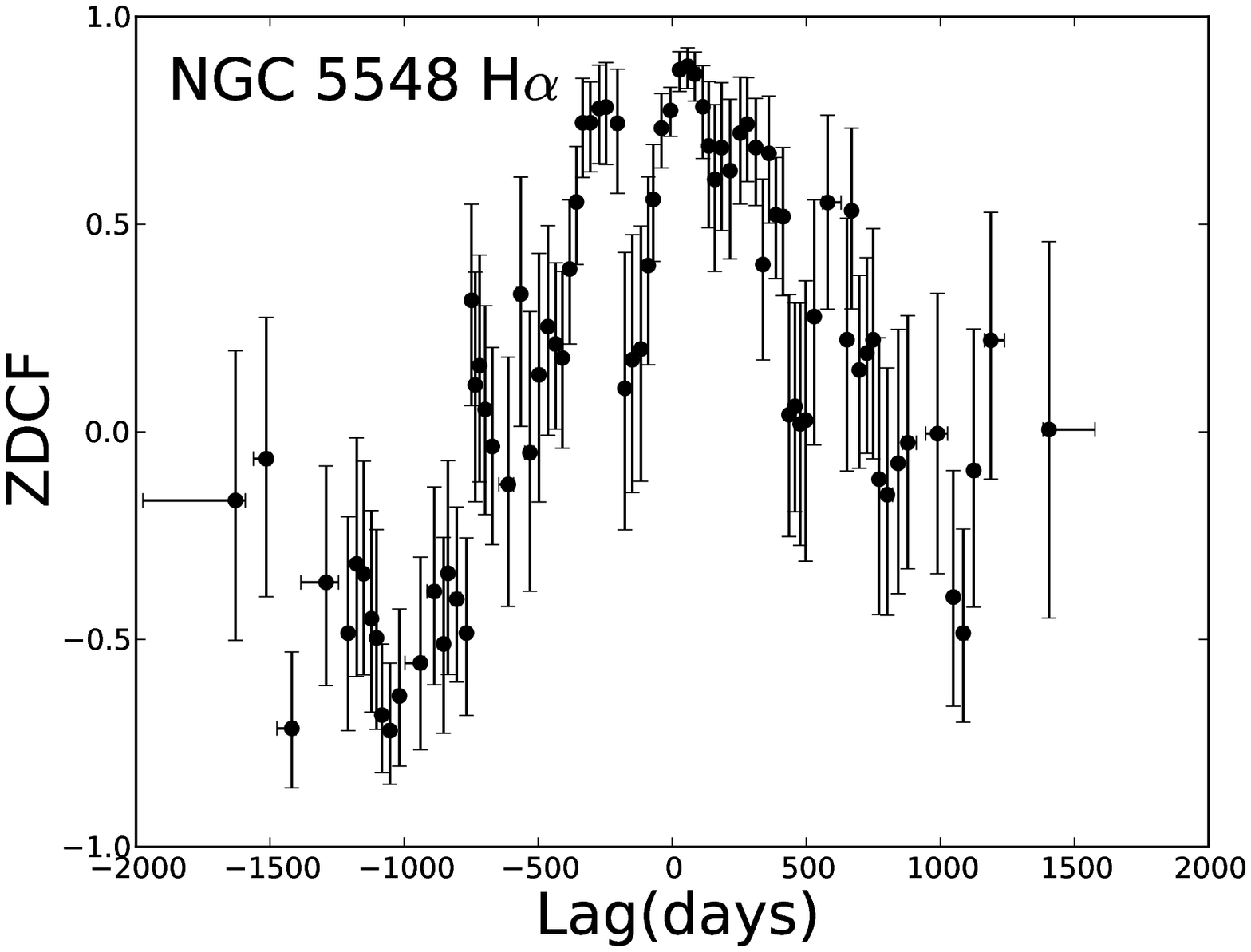}
\includegraphics[width = 6.cm]{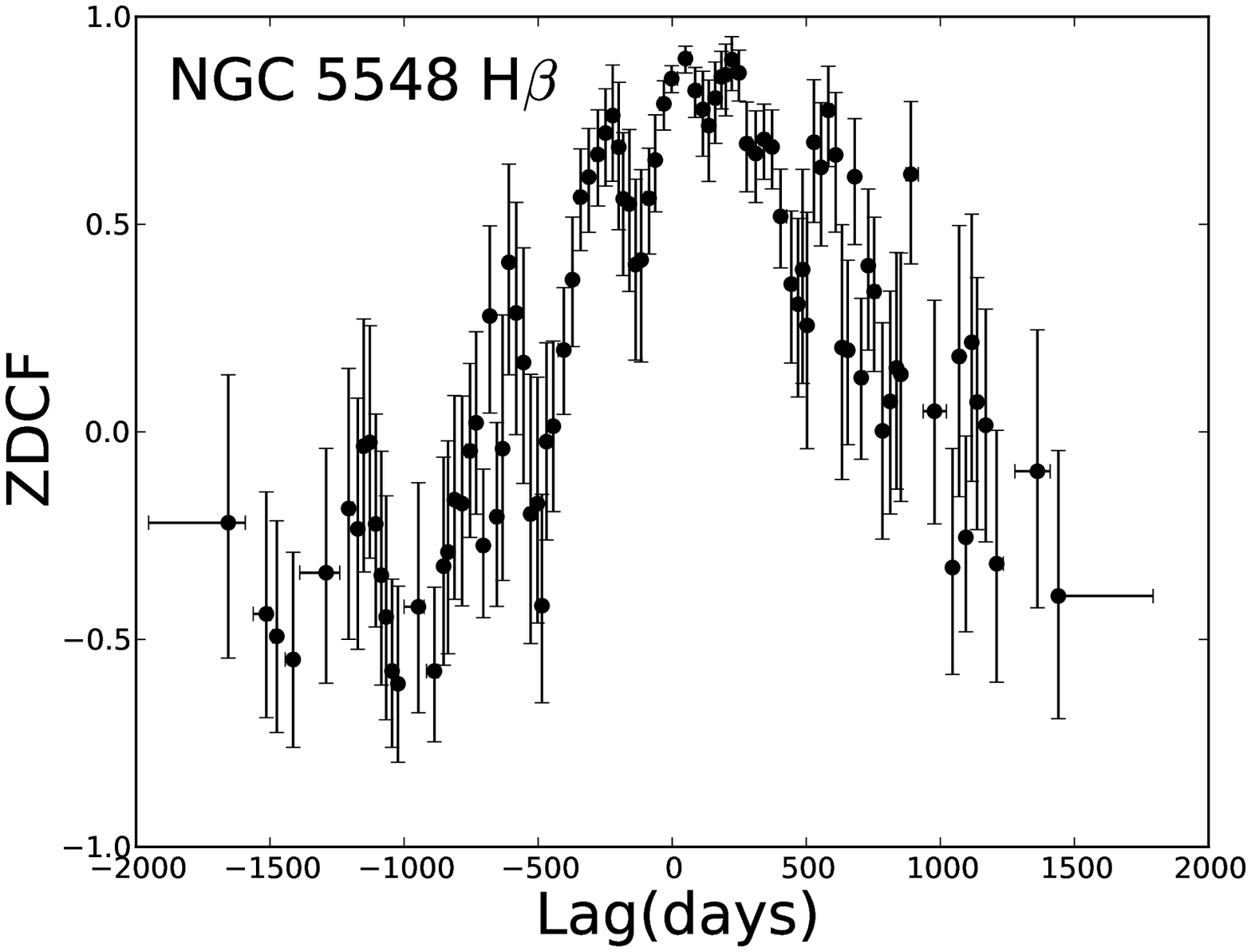}
\includegraphics[width = 6.cm]{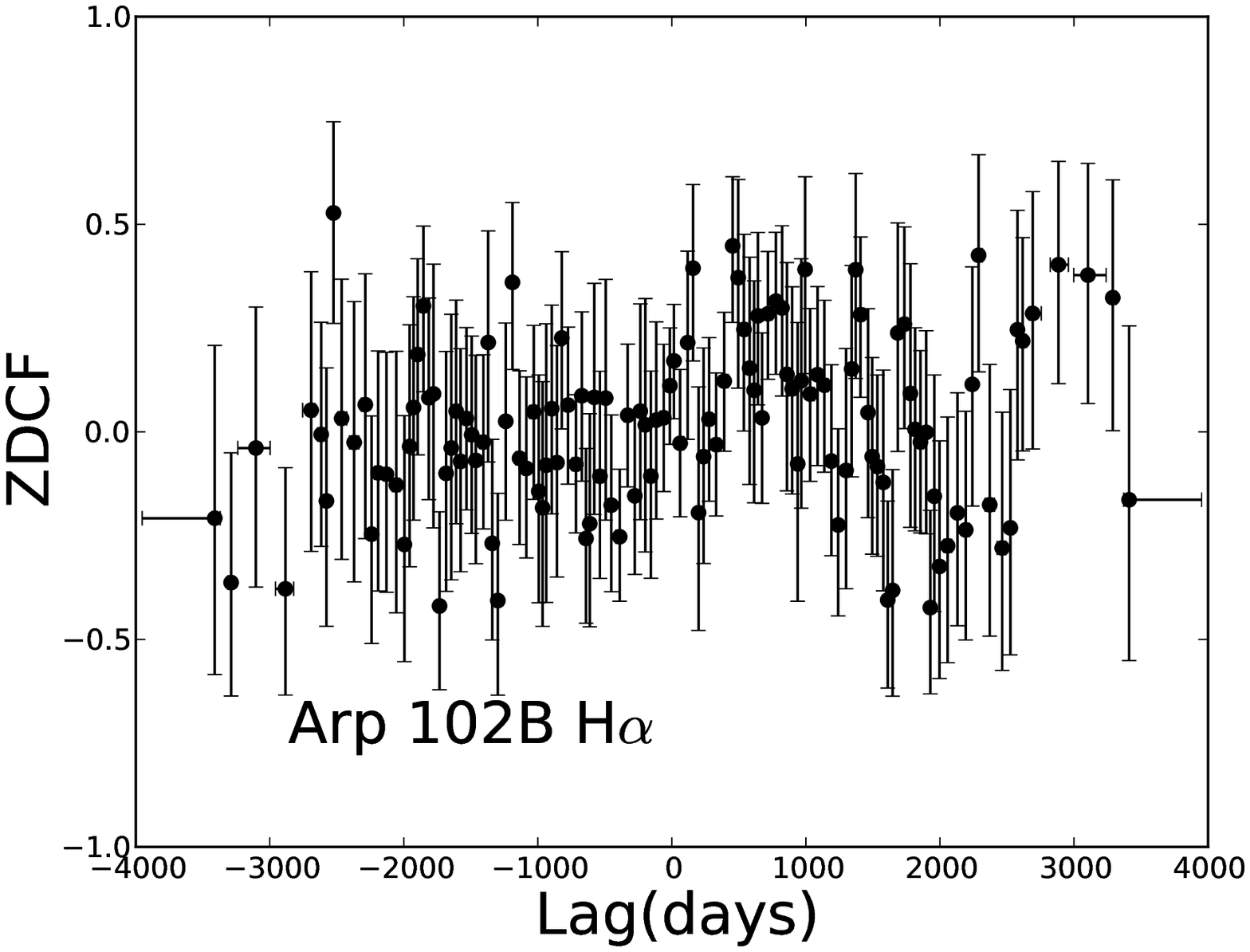}
\includegraphics[width = 6.cm]{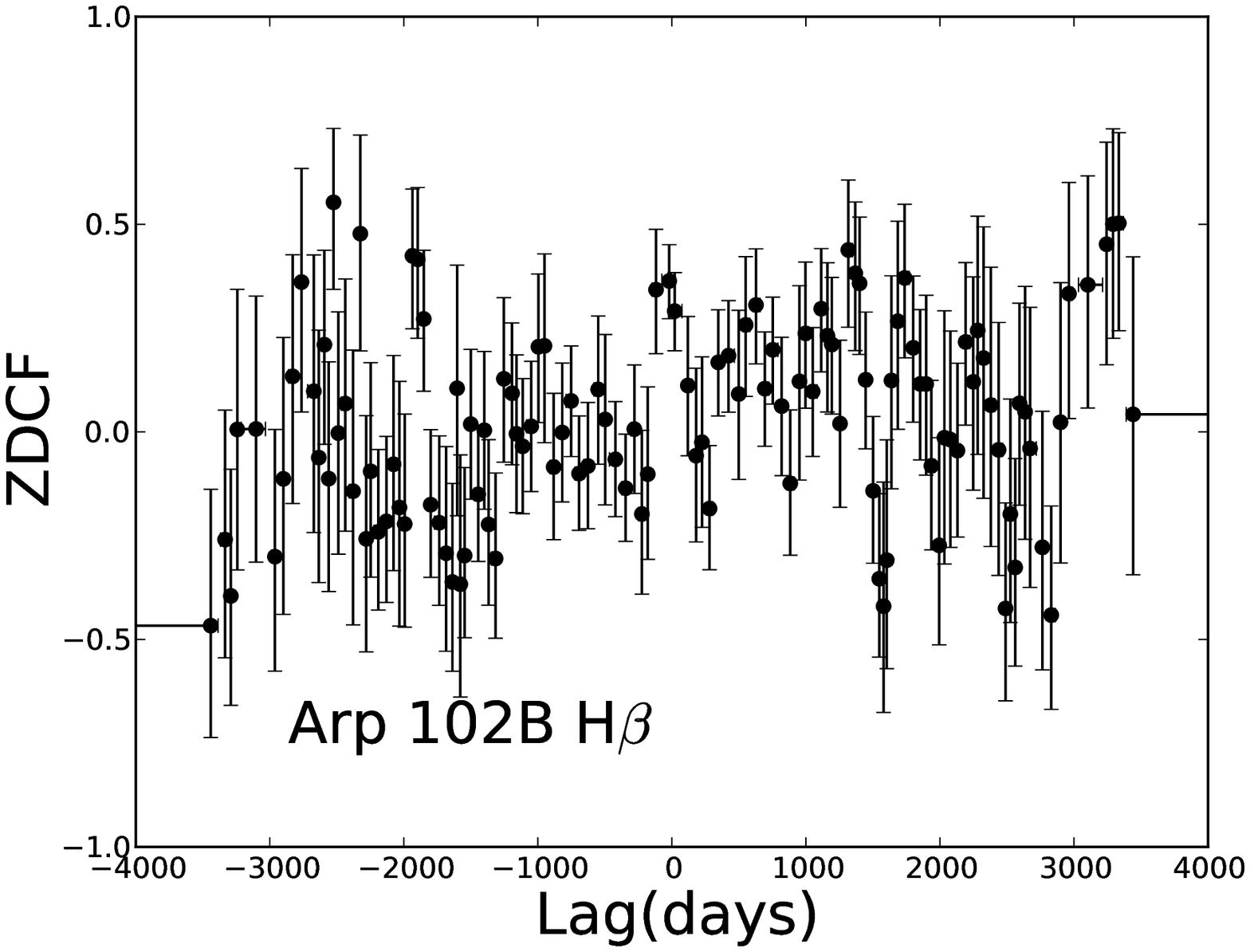}
\includegraphics[width = 6.cm]{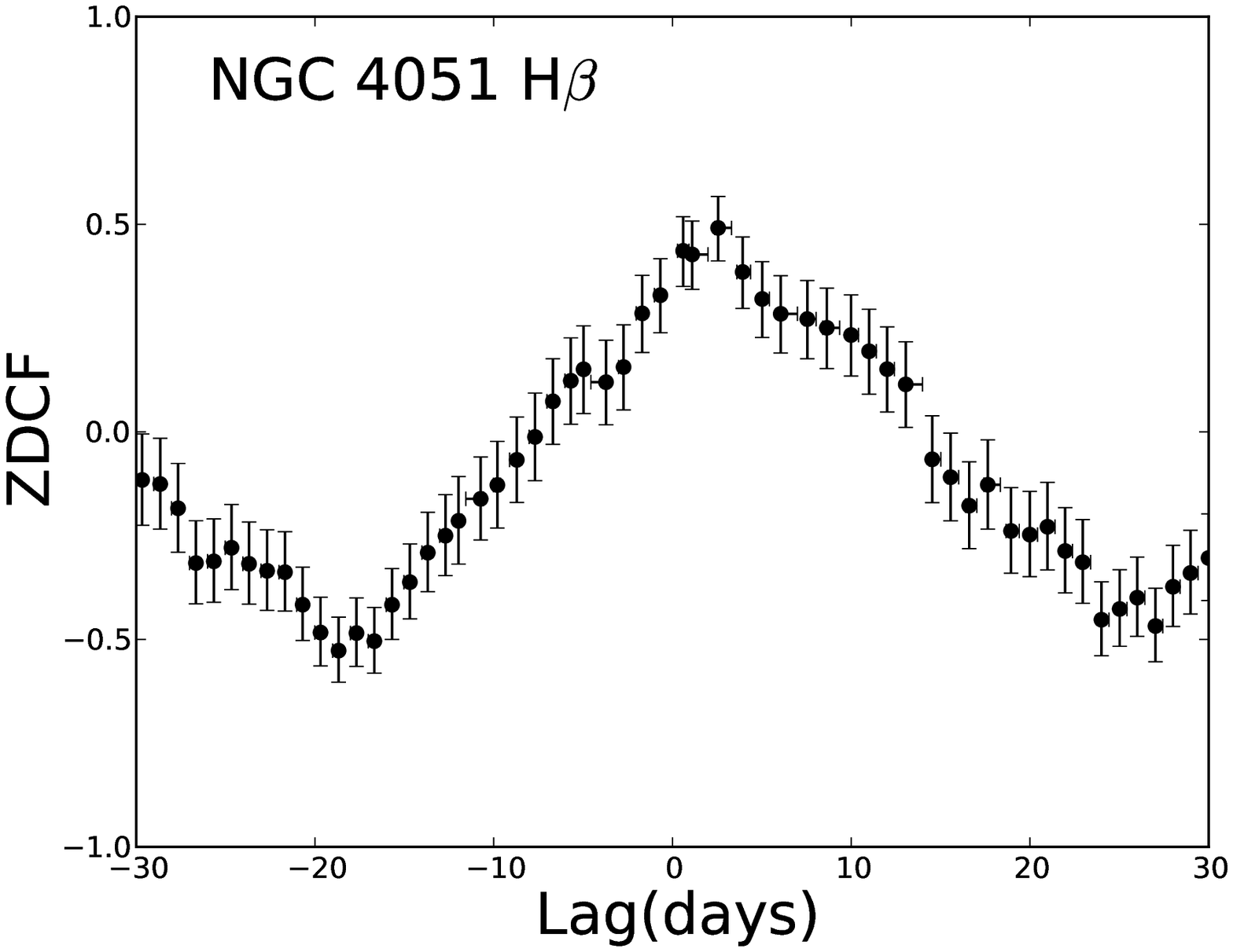}
\caption{The ZDCF analysis of object and emission line doneted on each plot. 
 The horizontal and vertical error bars correspond to $1\sigma$ uncertainties for a normal distribution.} 
\label{zdcf}
\end{center}
\end{figure}

   We apply the ZDCF method to the continuum and emission-line light curves of Arp 102B, 3C 390.3, NGC 5548, and NGC 4051,  
    considering the continuum measurements as the first time series and the line flux as the second one.
The results from the ZDCF analysis are presented in Fig. \ref{zdcf} and Table \ref{table2}. 
In Fig. 2 the ZDCF correlation coefficients as a function of time lag are shown, where the horizontal and vertical
 error bars represent the  $1 \sigma$   intervals in the time lags and correlation coefficients, respectively.
 Table \ref{table2} summarizes the following ZDCF parameters for each object: the time  lag  with the maximum likelihood coefficient 
  $\tau_{ML}$
 , the peak of ZDCF curve $r_{\rm max}$,  and the value of maximum likelihood parameter ML. 
 $ \bar{P_{cont}}$, $ \tilde{P_{cont}}$  are the mean and median 
  sampling periods for the continuum and
$ \bar{P_{line}}$,  $ \tilde{P_{line}}$  are the mean and median 
  sampling periods for the  line light curves, respectively, while N1 and N2 are
the total number of points in the corresponding light curve.  We use median vales 
for sampling period because median is much more resistant to outliers 
than is the mean. The time lag ($\tau_{ML}$) is basically the location 
of the peak of the correlation coefficient nearest to the zero lag, and it also has the largest maximum likelihood parameter. 
In the case of H${\alpha}$ line of  Arp 102B, in addition to the lag of around 15 days, there is also a lag of ${157.90}_{-46.33}^{+20.3}$ days, 
with large maximum {\bf likelihood of 0.87.  Therefore we also list} this result.
 In Table \ref{table2} the peak of maximum likelihood can be better determined in the case with more sampling points, 
as H$\alpha$ of Arp 102B and 3C 390.3
make it evident.



\begin{table}
\begin{center}
\caption{The ZDCF method results applied on the continuum (first time series) and emission line (second time series).
The positive lags mean that the second time series is delayed relative to the first.
In case of Arp 102B first two time lags are given. See text for details.}
\resizebox{12cm}{!}{%
  \begin{tabular}{llllllllllll}

\hline
\noalign{\smallskip}
Object& Time series 1&${ \bar{P_{cont}}}$  & $\tilde{P_{cont}}$   & N1 & Time series 2 & ${\bar{P_{line}}}$ &$\tilde{P_{line}}$   &N2 &$\tau_{ ML}$&$r_{ max}$ &ML \\
      & continuum    & (days)& (days) &  & emission line &(days)&  (days)& &    (days)       &  (days)  & \\

\hline
 
 Arp 102B&red&96.88&28.96&88& H$\alpha$&  94.66&28.96&90 &${15.0}_{-13.8}^{+24.4}$ &$0.20_{-0.14}^{+0.14}$ &0.79\\
& & && & && &&${157.1}_{-110.1}^{+20.3}$ &$0.39_{-0.22}^{+0.21}$ & 0.87\\
 &blue&73.08&20.98&116& H$\beta$&40.43&17.1&110 &${22.8}_{-20.9}^{+64.0}$&$0.23_{-0.1}^{+0.09}$&0.995 \\

\hline

 3C 390.3&5100 \AA \, &128.61&108.6&34& H$\alpha$&128.67&92.9&34 &${23.9}_{-10.5}^{+95.8}$&$0.90_{-0.06}^{+0.05}$&0.47  \\
 &5100 \AA \, &35.78&22.7&129& H$\beta$&35.78&22.7&129 &$94.5_{-48.0}^{+27.1}$&$0.94_{-0.03}^{+0.02}$&0.60 \\ 

\hline

 NGC 5548&5190 \AA \, &28.35&7.29&81& H$\alpha$&42.42&9.48&56 &$27.0_{-5.7}^{+14.4}$&$0.87_{-0.05}^{+0.04}$&0.99  \\
 &5190 \AA \, &28.35&7.29&81& H$\beta$&28.11&7.53&84 &$49.2_{-7.7}^{+18.6}$&$0.90_{-0.03}^{+0.03}$&0.997 \\

 \hline

 NGC 4051&5100 \AA \, &0.57&0.45&233& H$\beta$&1.08&1.0&108 &$2.6_{-1.1}^{+0.9}$&$0.49_{-0.08}^{+0.08}$&0.43 \\

 \hline
\end{tabular}
}
\label{table2}
\end{center}
\end{table}

\begin{table}
\begin{center}
\caption{The SPEAR method results based on fitting of the continuum and emission line. 
  The last three columns are giving the same results if the continuum and two lines 
(H$\alpha$ and H$\beta$) are fitted simultaneously. See text for details.}
\resizebox{10cm}{!}{%
  \begin{tabular}{llllllll}
\hline
Object& Line &$\tau_{SPEAR}$&$\sigma$&$\tau_{d}$ &$\tau_{SPEAR}^{\star}$& $\sigma^{\star}$&$\tau_{d}^{\star}$\\
                    
\hline
 Arp 102B& H$\alpha$&${23.8}_{-18.8}^{+27.5}$ &0.069 &45.15 &${31.5}_{-22.1}^{+47.3}$ &0.662 &430\\
 &H$\beta$&${47.6}_{-37.2}^{+57.2}$& 0.095&45.51 &${42.2}_{-27.5}^{+49.7}$& & \\
 \hline
 3C 390.3& H$\alpha$&${43.8}_{-34.8}^{+48.6}$&0.82  &1382.0 &${77.4}_{-59.0}^{+90.4}$&0.92  &1025 \\
 & H$\alpha$&${150.9}_{-143.8}^{+157.5}$&0.82  &1732.0 \\
 & H$\beta$&$76.9_{-74.7}^{+79.1}$& 0.94 &1147 &$76.0_{-55.0}^{+78.8}$& & \\ 
 \hline
 NGC 5548&H$\alpha$&$43.5_{-39.3}^{+47.7}$&  0.25    &476.38  &$45.5_{-39.9}^{+49.1}$&  0.273    &475  \\
 &H$\beta$&$45.4_{-43.0}^{+47.0}$&0.27         &415.45 &$43.4_{-38.4}^{+47.4}$&         & \\ 
 \hline
 NGC 4051&H$\beta$&$2.8_{-2.3}^{+3.1}$& 0.005&5.58 \\ 
 \hline
\end{tabular}
}
\label{table4}
\end{center}
\end{table}



\subsection{The SPEAR analysis}

  
 In order to perform SPEAR analysis , we build the continuum model to determine the DRW parameters of the continuum 
light curves for all four objects. 
Then, in order to measure the time lag between the continuum and H${\alpha}$ and H${\beta}$, 
SPEAR interpolates the continuum based on the posteriors derived, and then shifts, 
smooths, and scales each continuum light curve to compare to  H${\alpha}$ and H${\beta}$.  
   
\begin{figure}
\begin{center}
\subfloat[NGC 4051, top: cont., bottom: H${\beta}$ line]{\includegraphics[trim= 5 1 1 1, clip=true, width=0.5\textwidth]{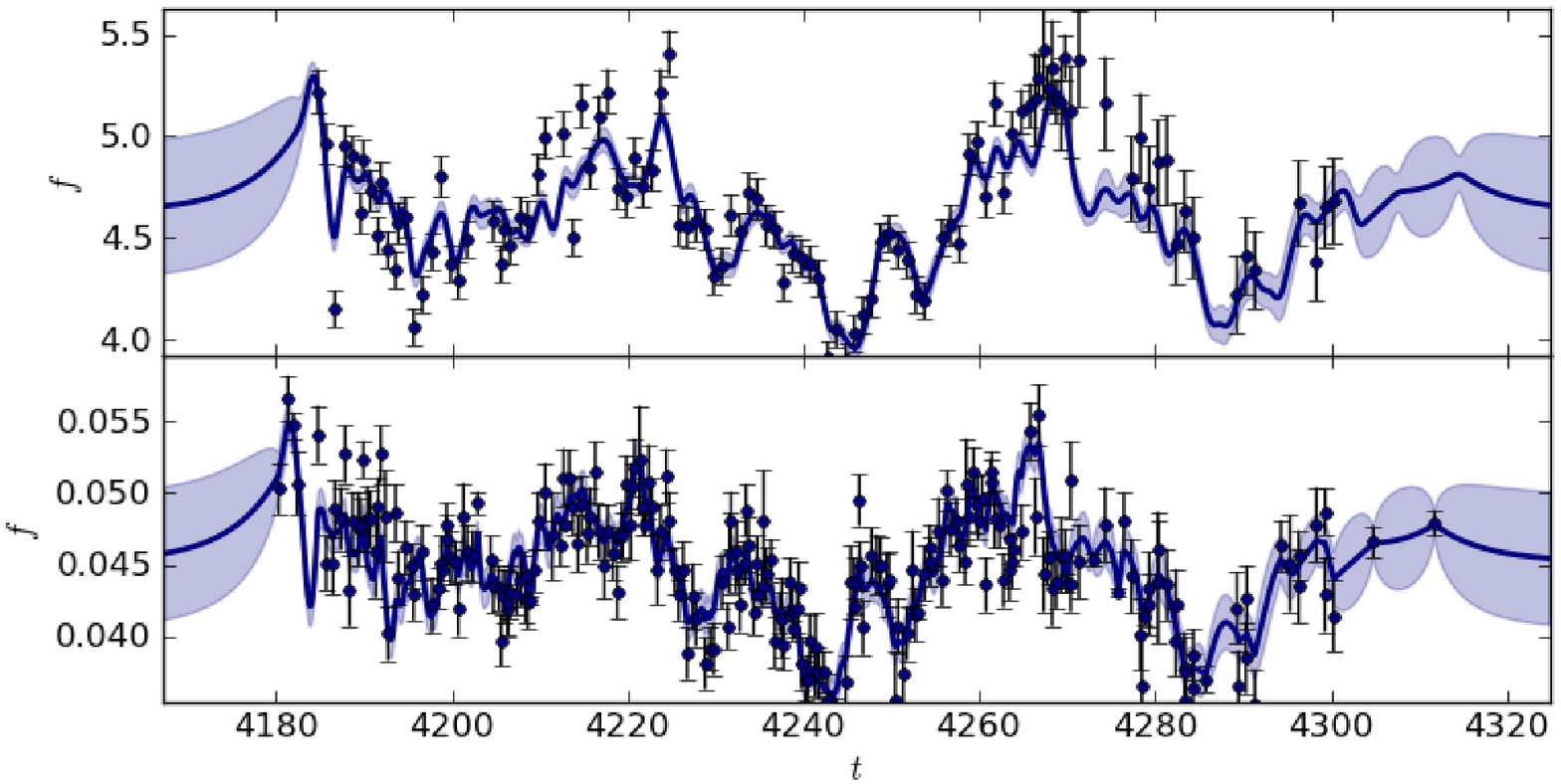}}
\subfloat[Arp 102B, top: cont., bottom: H${\alpha}$ line]{\includegraphics[trim= 5 1 1 1, clip=true,width=0.5\textwidth]{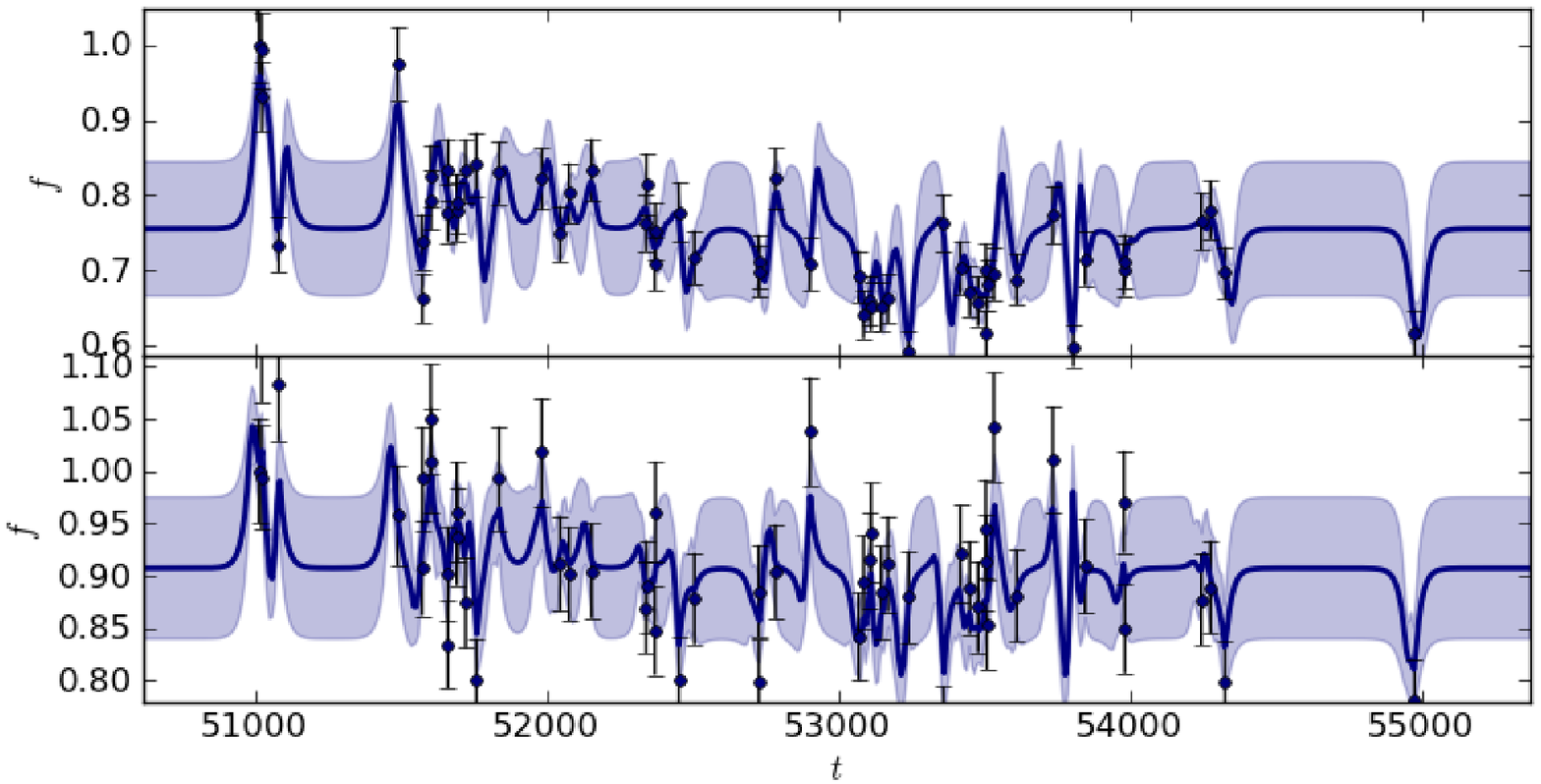}}\\
\subfloat[3C 390.3, top: cont., bottom: H${\alpha}$ line]{\includegraphics[trim= 5 1 1 1, clip=true,width=0.5\textwidth]{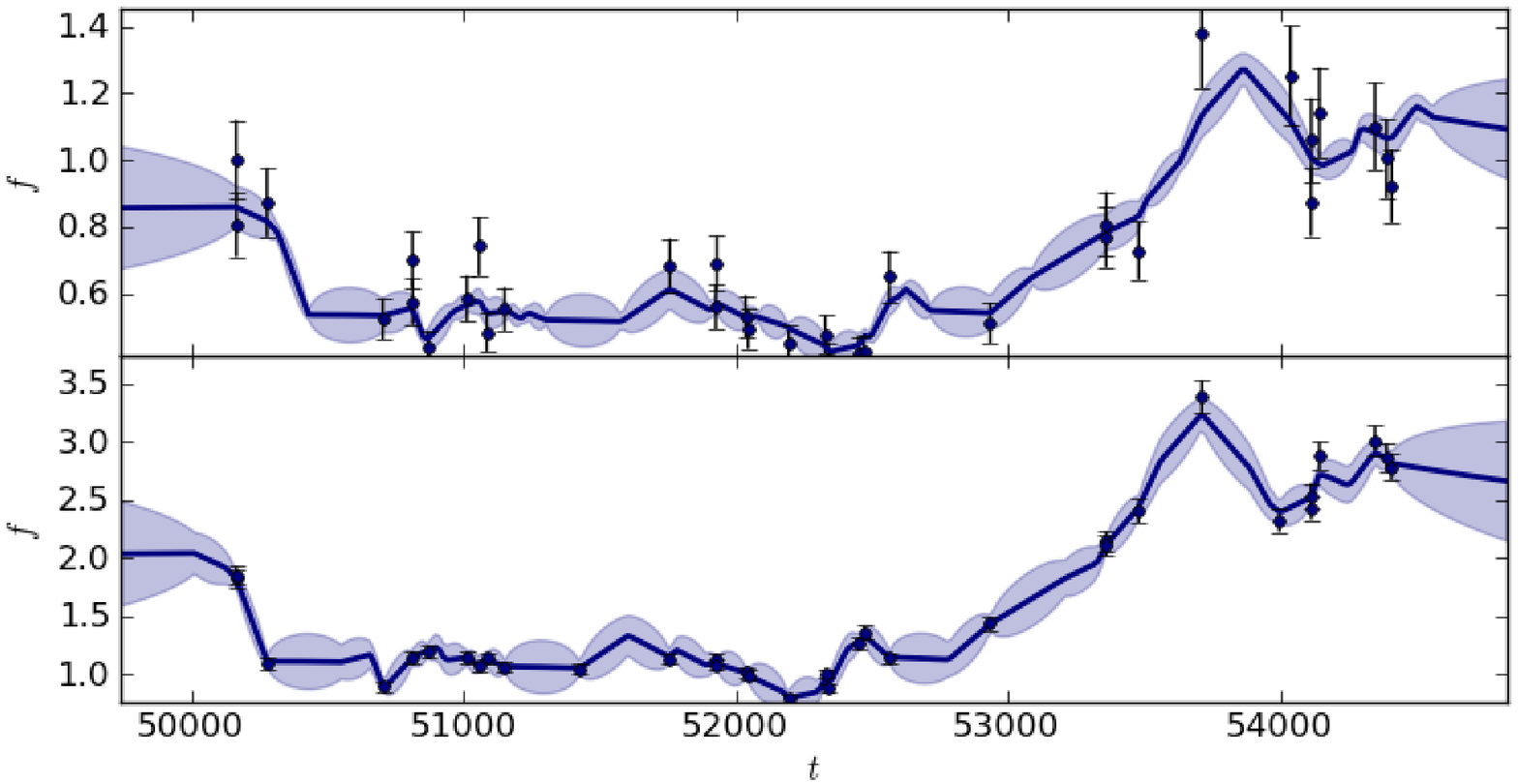}}
\subfloat[NGC 5548, top: cont., bottom: H${\beta}$ line]{\includegraphics[trim= 5 1 1 1, clip=true,width=0.5\textwidth]{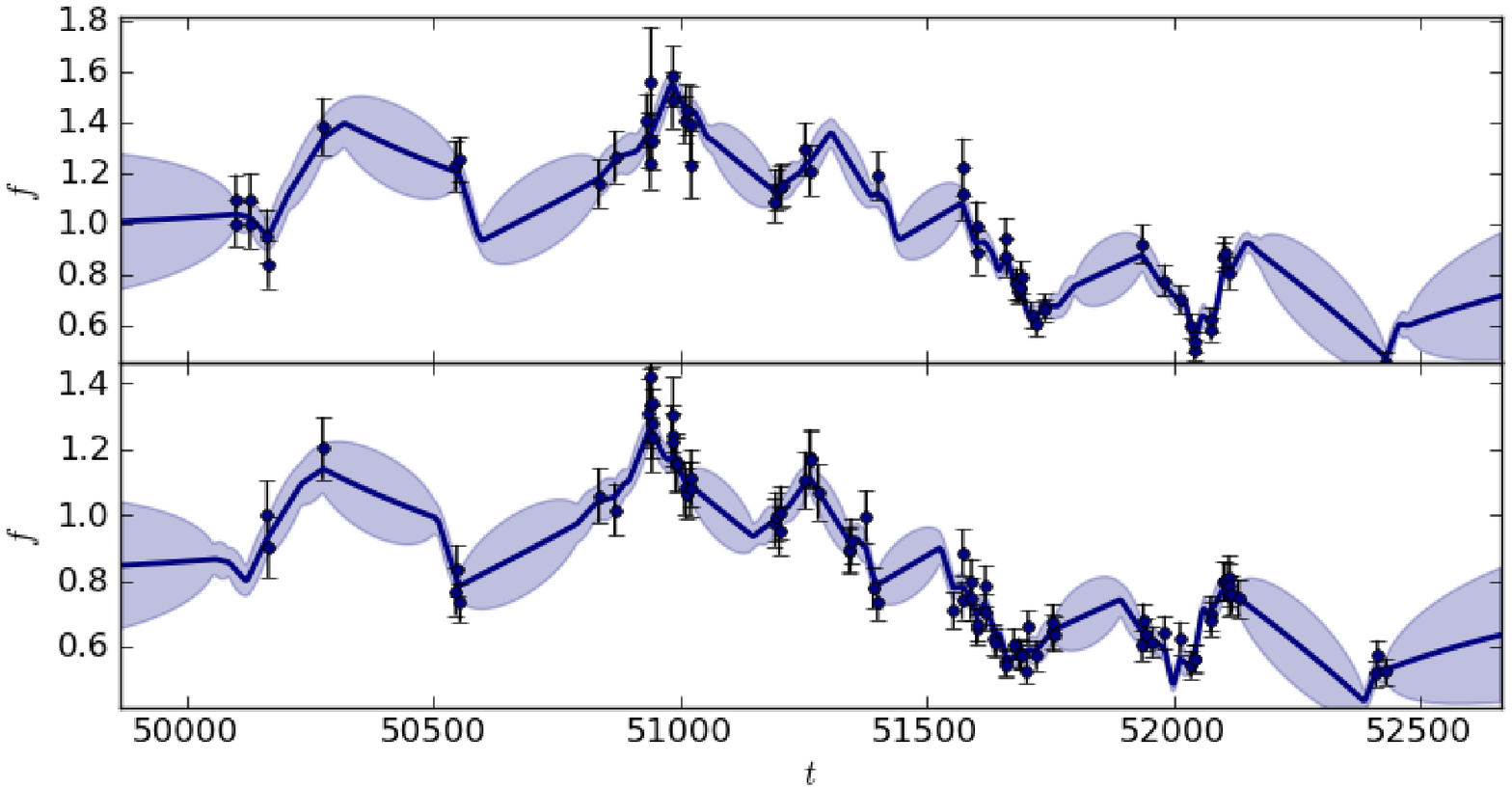}}\\
\caption{ The predicted mean light curves from the best-fit SPEAR model  of the continuum (upper plots)
and one emission line (bottom plots). 
The units in the case of NGC 4051 are $10^{-15} {\rm erg \, s^{-1} cm^{-2}}$ \AA \,$^{-1}$ for continuum, and $10^{-11} 
 {\rm erg \, s^{-1}  cm^{-2}}$ \AA \,$^{-1}$ for H${\beta}$, while in case of other objects the normalized light curves are given. 
 The black points with error bars show the observed light curves, the solid line shows the mean of the SPEAR light curve models 
 fit to the observed data, and the blue bands present the standard deviation \citep[see][]{Zu11}.}
\label{curves}   
\end{center}
\end{figure}
\begin{figure}
\begin{center}
\subfloat[NGC 4051,H$\beta$ line]{\includegraphics[height=2.0in,width = 2.5in]{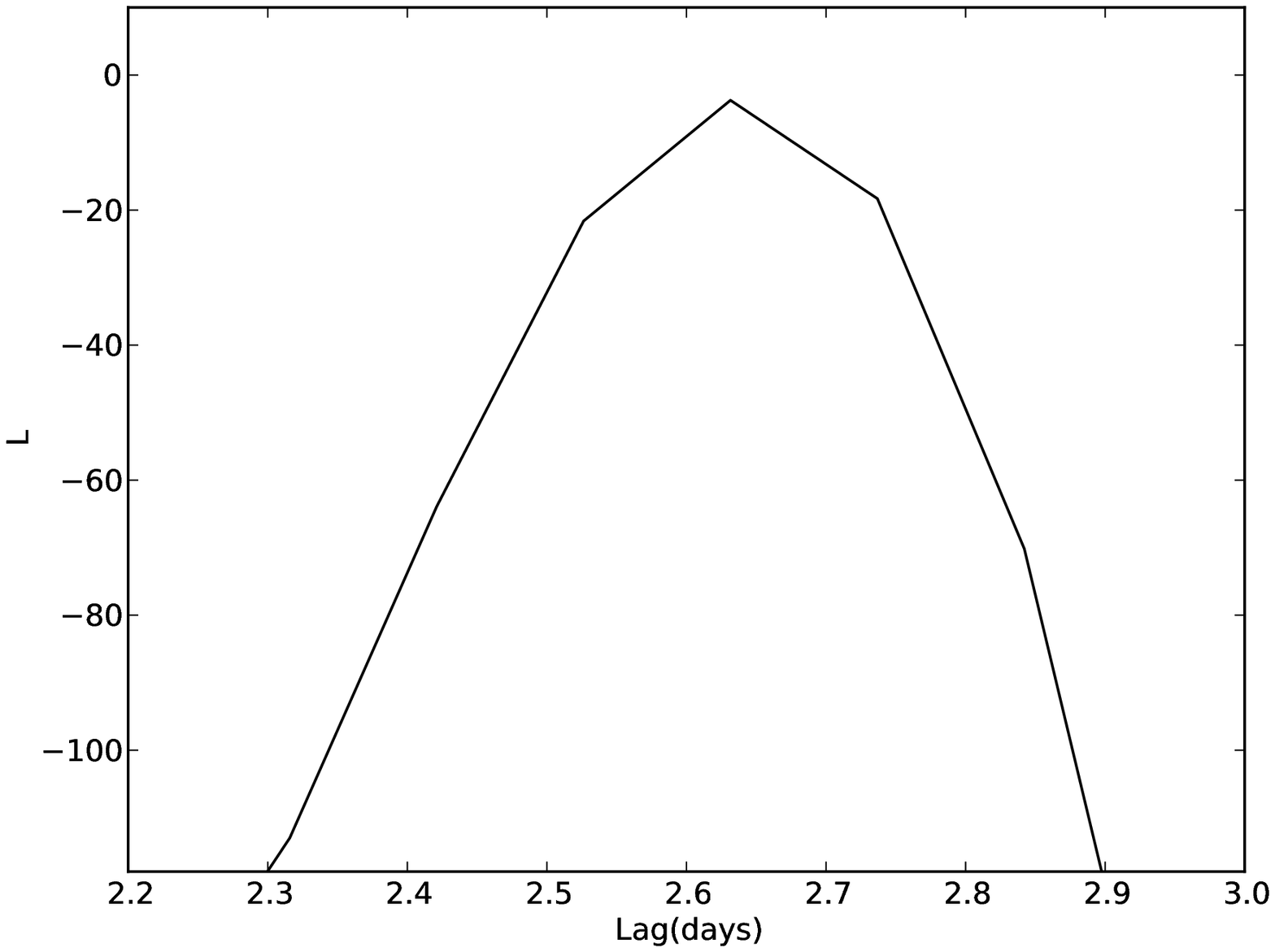}}
\subfloat[Arp 102B,H$\alpha$ line]{\includegraphics[height=2.0in,width = 2.5in]{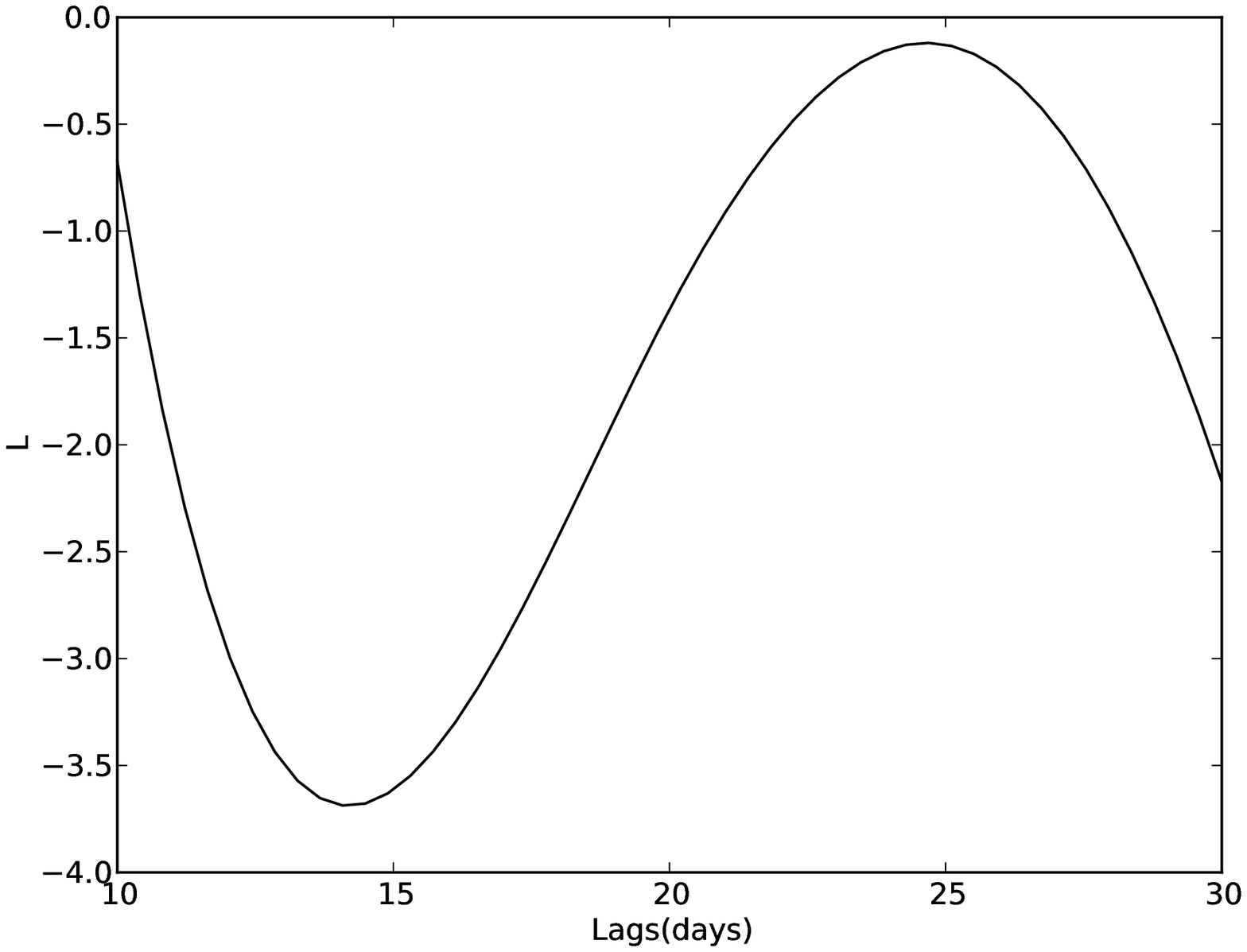}}\\
\subfloat[NGC 5548,H$\alpha$ line]{\includegraphics[height=2.0in,width = 2.5in]{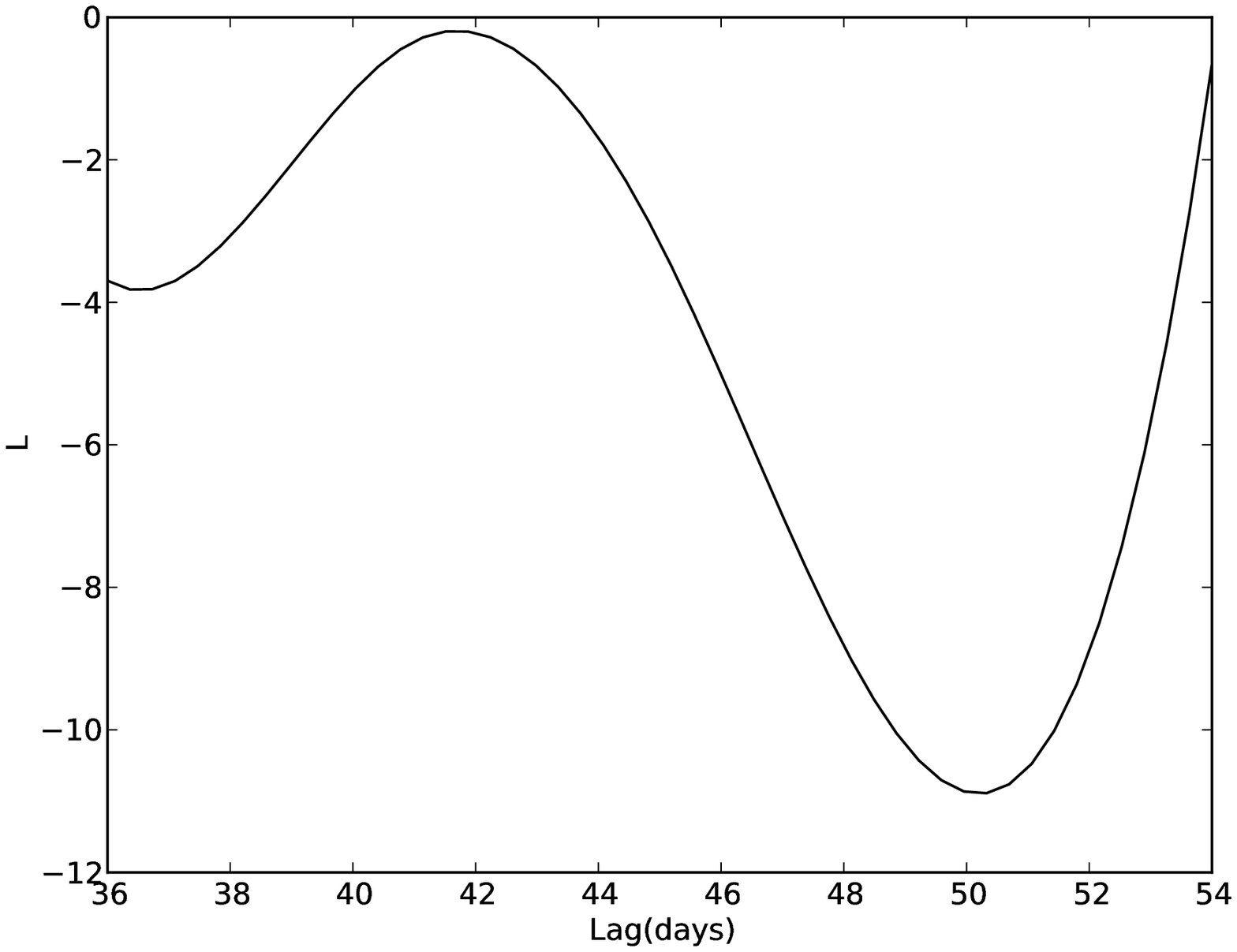}}
\subfloat[3C 390.3,H$\alpha$ line]{\includegraphics[height=2.0in,width = 2.5in]{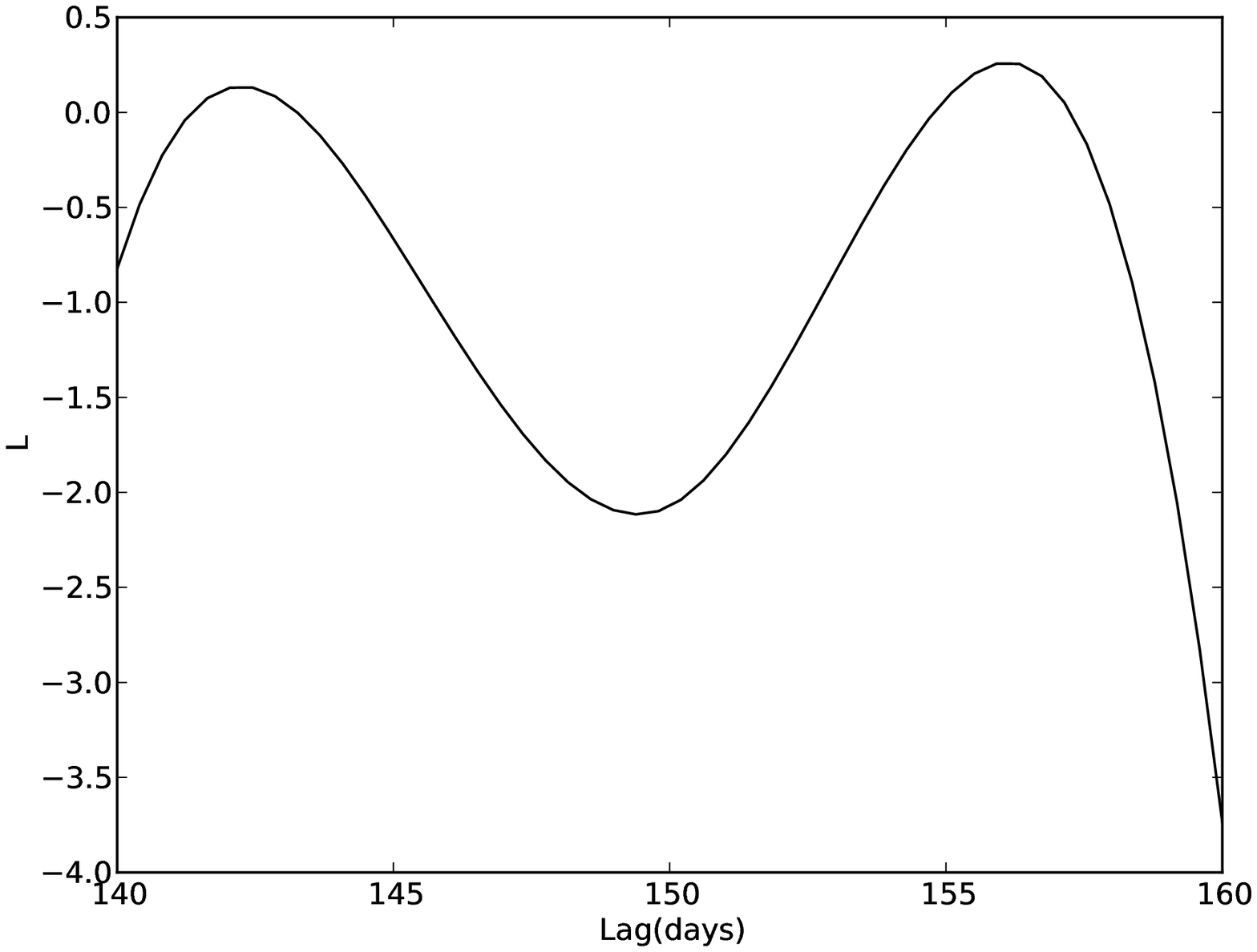}} 
\caption{The  ratio of log-likelihood functions  from the SPEAR analyses. The y-axis shows the SPEAR likelihood 
ratios for {\bf  continuum and single emission line as labeled}. The log-likelihood 
ratio is negative since it is calculated for a simple {\bf vs simple hypothesis.
The number  of parameters }in the models are the same,
  while the range of lag is varied: wider range  and narrower range around obtained lag. The ranges of time lag $\tau$ used in
  fitting procedure are enclosing time lags obtained from continuum+one line and unique 
solution obtained from continuum+two line fitting procedure. The unique solution obtained from continuum+two line fitting  eliminates spurious lags.
The ratio of log likelihoods can not pass the narrower range of lags used in testing hypothesis.  
 In the case of 3C 390.3 is given ratio  log-likelihoods for range of time lag 
enclosing the secondary solution from continuum+one line fitting SPEAR's procedure.} 
\label{so}
\end{center}
\end{figure}

Examples of the mean light curve models of the continuum and emission line, obtained by the SPEAR fitting, 
are presented in Fig. \ref{curves}. Fig. \ref{so} gives $\Delta \chi ^{2}$ from Eq. (5) between models.  
The time lag values $\tau_{SPEAR}$, the updated posteriors of the damping time scale $\tau_{d}$
and the variability amplitude $\sigma$
are given in Table \ref{table4}.  The values obtained from fitting the continuum 
and a single line are given  in the first set of columns. The results for fitting the continuum 
and both H$\alpha$ and H$\beta$ line simultaneously 
with SPEAR method are also given in  Table \ref{table4} (last three columns). The  values of $\tau_{\rm SPEAR}$ 
   for  H$\alpha$ and H$\beta$ obtained from simultaneous fitting procedure are 
slightly different.
{\bf The  contrast arises from differences in sampling periods and number of data points
 of continuum  curves  around H$\alpha$ and H$\beta$ lines respectively 
  (see Table \ref{table2}). E. g,  SPEAR  simultaneous procedure 
   has been applied on  two types of set of curves.  The first model consisted of  the  continuum around  H$\alpha$ line, 
   the 
    H$\alpha$ line (as the first line) and H$\beta$ line (as the second), while the second model consisted of
      the  continuum around H$\beta$ line, 
    H$\beta$ line (as the first line) and H$\alpha$ (as the second line).}


The slow (or lack of) convergence and existence of multiple peaks of the log-likelihood function
 (Fig. \ref{so}) could be influenced by:  sampling characteristics of data set \citep{Zu11},
  the relationship among variables of set of observations as well as their values even in the case of
   well sampled data \citep{Gri12}. In the case of our  data set of NGC 5548,  H$\alpha$ line’s average sampling period 
is twice larger than both the continuum and H$\beta$ line,  while median sampling periods are almost similar. 
The average sampling periods of Arp 102B H$\alpha$ line and corresponding 
continuum are quite large (about 100 days, almost 3 times larger than median sampling period,
 which indicates skewness of its data distribution).
 
 In the case of H$\alpha$ and continuum  of 3C 390.3,
  the both the median and average sampling periods are the largest {\bf among the sample of 
  AGNs. Its log-likelihood function} shows two peaks and slow convergence. 
The curves are actually  modeled as a Gaussian process,
  while in  principle there could be deviations of light curves from Gaussian process infecting the fitting parameters. 
The large multi-year gaps in the data would result in weaker constraints on the final model fit, since the covariance matrix is 
  calculated for given differences between  observational times. 
Also, the parameters $\tau_{d}$ and $\sigma_{d}$  are found to be correlated with the
 physical properties of accretion disks, including optical luminosities, 
 Eddington ratios, and black hole masses, which also could affecting fitting results of the SPEAR method (DRW is actually phenomenological model).

\section{Discussion}
 
{\bf For comparison, we summarize the time lag measurements} given in the literature
 for these  4 objects in Table \ref{history}. {\bf Our results from both} methods are comparable (not congruent) with the earlier estimates, 
but some differences could be found.

The largest time lags are for 3C 390.3 and NGC 
5548 (see Table \ref{history}). For 3C 390.3 the longest monitoring campaign  was undertaken
 by \citet{Sh10} and \citet{DP12}. Analyzing the discrepancy between the time lags 
 obtained from these two campaigns, \citet{DP12} found that the properties of the continuum strength 
 variations have a significant effect on the time delay obtained using CCF. Also, \citet{SP02, SK11} noted that
 the width of the auto-correlation function (ACF)  of the continuum light curve which is covering several years
  up to more than a decade is much broader than the ACF of a shorter campaign,
 and that this will result in a longer time delay. 
   Since the CCF is the convolution of the transfer function with 
ACF of the continuum, and  the 
ACF is a symmetric function, asymmetric properties of the transfer function (e.g. the position of the 
peak) must be reflected in a similar manner in the CCF   
 \cite[{\bf e. g. Mrk 50;}][]{Pan12}. 
However, it is  also possible that changes in the 
 measured cross-correlation lag arise from changes 
 in the continuum ACF rather than in the delay-map 
 (see \citet{RP90}, \citet{PRF92a}, \citet{PRF92b}, \citet{W99}). 
  If the continuum variations become slower, 
  a sharp peak at low time-delay 
  in the delay distribution will be blurred by the broader 
  ACF{\bf ,}  and the peak of the CCF will be shifted to larger delays {\bf \citep{NM90}}. 
 Thus, the lag measured by cross-correlation analysis depends not only on the delay distribution, transfer function,  
   but also on the characteristics ACF of the continuum variations.
 The monitoring campaining of \citet{DP12} lasted for 80 days and they could not measure longer delays. 
 On the other hand, wider temporal sampling of the long monitoring campainings could affect results too.  
 For NGC 5548,  our calculations produced larger values for H$\beta$ time lags, while
 for NGC 4051 we could see that time lag obtained by \citet{DP06} is about 3 
 times smaller than the value obtained by \citet{Pet00}, and our calculation 
 confirms value of about 2.5 days.

\begin{table}
\begin{center}
\caption{Time lag measurements  for Arp 102B, 3C 390.3, NGC 5548, and NGC 4051.}
\resizebox{14cm}{!}{%
  \begin{tabular}{llllll}
  \hline
Object& Continuum waveband& Line  &  $\tau$ & Method used & References \\
      &  (in \AA \,)         &       &  (days) & & \\
\hline

Arp 102B &cnt 6368-6412& H$\alpha$ &$13_{-105}^{14}$&ICCF&\citet{S00}  \\
& cnt 6356-6406& H$\alpha$ &$15_{-13.8}^{15.7}-17_{-14.3}^{14.3}$&ZDCF&\citet{Sh13}  \\
               &           &        &$23_{-30.2}^{16.8}  $ & SPEAR  &  \citet{Sh13} \\

& cnt 6356-6406& H$\alpha$&$ 15_{-13.8}^{24}$,$24_{-18.8}^{27.5}$& ZDCF,SPEAR & this work \\

& cnt 5200-5250& H$\beta$ &$11_{-9.8}^{19.3}-21_{-19}^{54.3}$&ZDCF&\citet{Sh13}  \\

               &           &        &$37_{-47.5}^{19.7} $  & SPEAR  &  \citet{Sh13} \\      

& cnt 5200-5250& H$\beta$& $23_{-20.9}^{64}$,$48_{-37}^{57}$& ZDCF,SPEAR & this work \\

\hline

 3C 390.3&cnt 4400-9000& H$\alpha$&$20\pm8$&ICCF+ZDCF+DCF&\citet{D98}  \\
 & cnt 6484-6608& H$\alpha$ &$162_{-15}^{32}$&ICCF&\citet{SP02}  \\ 
 
  & cnt 5369-5399& H$\alpha$ &$23_{-7}^{5}$&ZDCF+ICCF&\citet{Sh10}  \\
    & cnt 6200& H$\alpha$ &$174\pm16$ &ICCF&\citet{SK11}  \\
  
   & cnt 5100& H$\alpha$ &$56.3_{-6.6}^{2.4}$&ICCF&\citet{DP12}  \\
     &          &        &$52.5_{-0.6}^{0.7}$ & SPEAR   & \citet{DP12}  \\   
 
  & cnt 5369-5399& H$\alpha$& $24_{-10.5}^{95.8}$,$44_{-35}^{49}$& ZDCF,SPEAR & this work \\ 
  
  &cnt 4400-9000& H$\beta$&$20\pm8$&ICCF+ZDCF+DCF&\citet{D98}  \\  
  & cnt 4400-9000& H$\beta$ &$24.2_{-8.4}^{6.7}$&photoiozination&\citet{WPM99}  \\
 & cnt 5000-5006 & H$\beta$ & $ 50_{-10}^{100}$  &ICCF&\citet{P04}  \\   
 & cnt 5370-5420& H$\beta$ &$100_{-97}^{101}$&ICCF&\citet{Sh01}  \\
  & cnt 5360-5425& H$\beta$ &$89_{-10}^{12}$&ICCF&\citet{SP02}  \\  
  & cnt 5369-5399& H$\beta$ &$96_{-47}^{28}$&ZDCF+ICCF&\citet{Sh10}  \\
   & cnt 5100 & H$\beta$ &$94\pm16  $&ICCF&\citet{SK11}  \\ 
     & cnt 5100& H$\beta$ &$44.3_{-3.3}^{3}$&ICCF&\citet{DP12}  \\
  
     &          &        &$47.9_{-4.2}^{2.4}$ & SPEAR   & \citet{DP12}  \\ 
          & cnt 5369-5399& H$\beta$&$ 95_{-48}^{27}$,$77_{-75}^{79}$& ZDCF,SPEAR & this work \\   
          &cnt 4400-9000& H$\gamma$&$20\pm 8$&ICCF+ZDCF+DCF&\citet{D98}  \\   
          & cnt 5100& H$\gamma$ &$58.1_{-6.1}^{4.3}$&ICCF&\citet{DP12}  \\
     &          &        &$32.1\pm 17.3$ & SPEAR   & \citet{DP12}  \\         
       

            & cnt 5100& HeII &$22.3_{-3.8}^{6.5}$&ICCF&\citet{DP12}  \\
  
    &          &        &$36\pm 5.2$ & SPEAR   & \citet{DP12}  \\

& cnt 1330-1470& Ly$\alpha$&$ 69_{-19}^{50}-108_{-21}^{15}$      &ICCF+DCF&\citet{WW97}\\
             &     CIV      &        &$49_{-17}^{13}-63_{-20}^{57}$   &   &  \citet{WW97}\\

       &cnt 1340-1400     & Ly$\alpha$ &$ 34\pm17$ &ICCF+DCF&\citet{OB98}  \\

        &     CIV      &        &$66\pm35$  &   &\citet{OB98}\\ 
            
 
  
     
  
  
 \hline

NGC 5548   & cnt 6300-6350& H$\alpha$ &$11.02_{-1.15}^{1.27}$& ICCF&\citet{Bentz10n}  \\

& cnt 5190& H$\alpha$&$ 27_{-6}^{14}$,$43_{-39}^{48}$& ZDCF,SPEAR & this work \\

& cnt 5185-5195& H$\beta$ &$6_{-4.08}^{4.44}-26.4_{-2.63}^{4.67}$&ICCF&\citet{Pet02}  \\
 & cnt 5170-5200& H$\beta$ &$6.5_{-2.5}^{2.5}$& ICCF+DCF&\citet{BD07}  \\
& cnt 4800-4830& H$\beta$ &$18_{-5}^{5.8}-25_{-1.5}^{1.4}$&ICCF&\citet{SD07}  \\
& cnt 5150-5200& H$\beta$ &$4.25_{-1.33}^{0.88}$& ICCF&\citet{Bentz09}  \\
& cnt 6300-6350& H$\beta$ &$4.25_{-1.33}^{0.88}$& ICCF&\citet{Bentz10n}  \\
& cnt 5190& H$\beta$& $49_{-7.7}^{19},45_{-43}^{47}$& ZDCF,SPEAR & this work \\
& cnt 6300-6350& H$\gamma$ &$1.25_{-2.34}^{1.86}$& ICCF&\citet{Bentz10n}  \\

 
 \hline
NGC 4051  & cnt 5100-6800& H$\alpha$ &$3.1\pm1.6$&DCF&\citet{ShU03}  \\

& cnt 5090-5120& H$\beta$ &$6\pm 3$&ICCF+DCF&\citet{Pet00}  \\

  & cnt 5100-6800& H$\beta$ &$2\pm2.3$&DCF&\citet{ShU03}  \\

& cnt 5090-5130& H$\beta$ &$1.87_{-0.5}^{0.54}$&ICCF+DCF&\citet{D09c}  \\
  & cnt 5100& H$\beta$ &$3.5_{-0.5}^{12}$&ICCF&\citet{Y13}  \\
    & cnt 5190& H$\beta$&$ 3_{-1.1}^{0.9},3_{-2.3}^{3.1}$& ZDCF,SPEAR & this work \\  
\hline
 \end{tabular}
 }
\label{history}
\end{center}
\end{table}


  In order to test the statistical relationship between the ZDCF and SPEAR results, 
  we calculated the Pearson correlation coefficient $r$ and the p-value between different sets of time 
lags,  which confirmed that results obtained from H$\alpha$ of 3C 390.3 are an outlier.
Such behavour of H$\alpha$ line of this object could have roots in bad median and average sampling period 
and very broad the ACF of the continuum and the line itself.
 
The secondary (spurious) time lags are obtained for H$\alpha$ lines of  Arp 102B and
3C 390.3. These spurious lags could arise due to the fact that the H$\alpha$ line 
and continuum of both objects {\bf have large sampling period. 
In the case of  3C 390.3 average and median sampling periods are larger than  100 days and in
the case of Arp 102B average sampling period is 100 days. The median value  is three times smaller, 
indicating skewness of data distribution.
Other two objects} have much smaller
sampling periods. The ACF
of continuum and H$\alpha$ of 3C 390.3 are very broad, which
can also lead to spurious large values of CCF  \citep[see][]{W99}. As for 
the secondary lag of Arp 102B, we noticed that the autocorrelation functions of
continuum and H$\alpha$ line are noisy and similar to each other (it
could be a consequence of irregular sampling and physical nature of light
curves itself).

Since we have spurious time lags,  we defined both sets with and without these values.
We found the strongest correlation ($r \sim 0.9$) between the ZDCF and one-line SPEAR fitting 
time lags (when spurious lags of H$\alpha$  of Arp 102B and 3C 390.3 are excluded
in ZDCF and SPEAR sets, respectively). Also, we found a weaker correlation 
between the sets of ZDCF time lags (without spurious lag included, the unique solution 
from SPEAR's continuum + two line  fitting procedure eliminates second peak) and  
two-line SPEAR fitting time lags. This is not surprising since when we fit both
lines simultaneously, the light curve together with its determined lag
adds extra information to the continuum light curve, and thus better
constrains the another light curve's lag. Both correlations  are
suggesting that these sets of time lags are reliable and linearly
dependent. All other combinations which include spurious time lags of H$\alpha$
Arp 102B and 3C 390.3 in ZDCF and SPEAR sets are linearly
uncorrelated, which allowed us to discard these spurious peaks.

\begin{figure}
\begin{center}
\includegraphics[width = 8.cm]{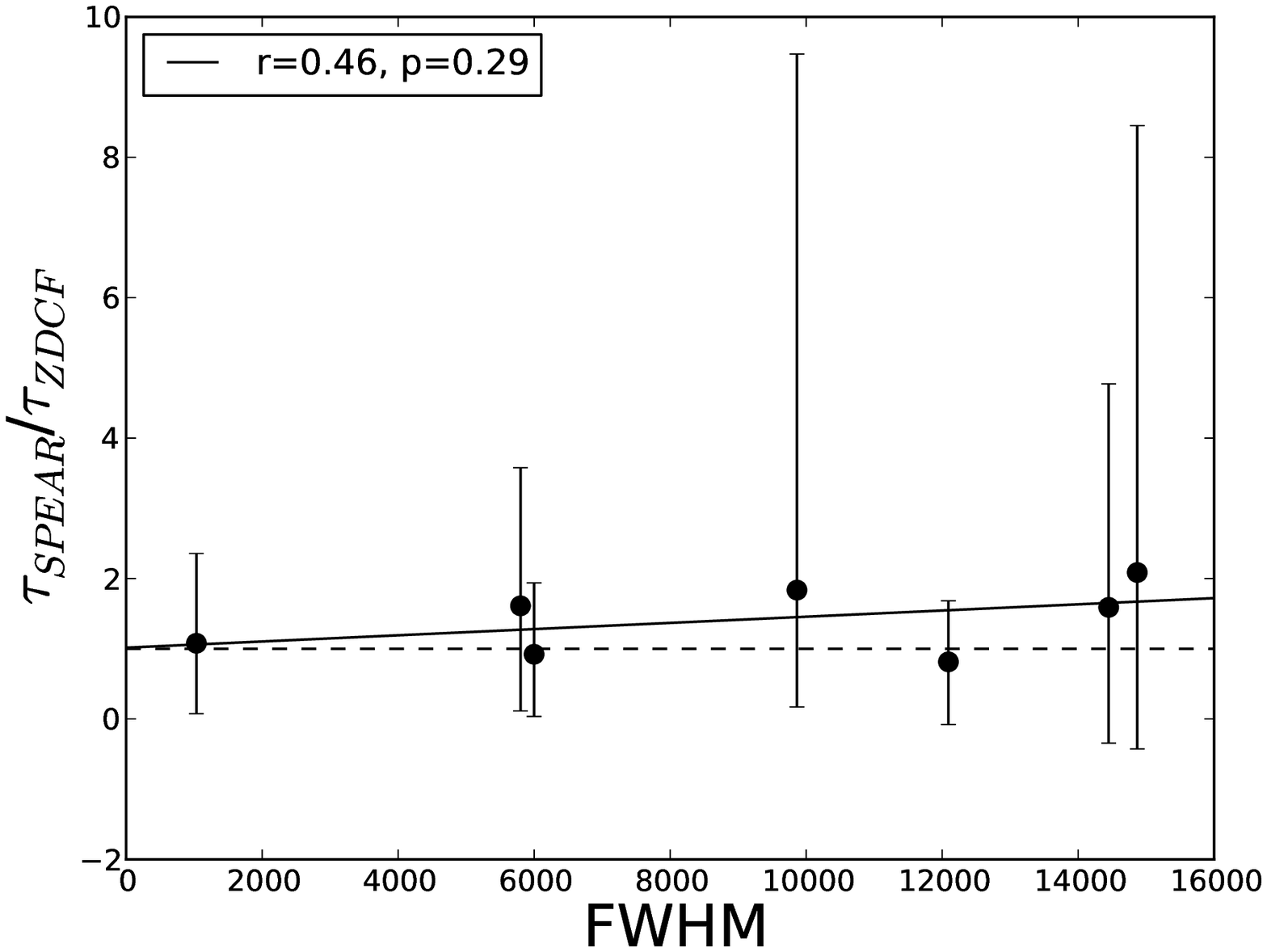}
\includegraphics[width = 8.cm]{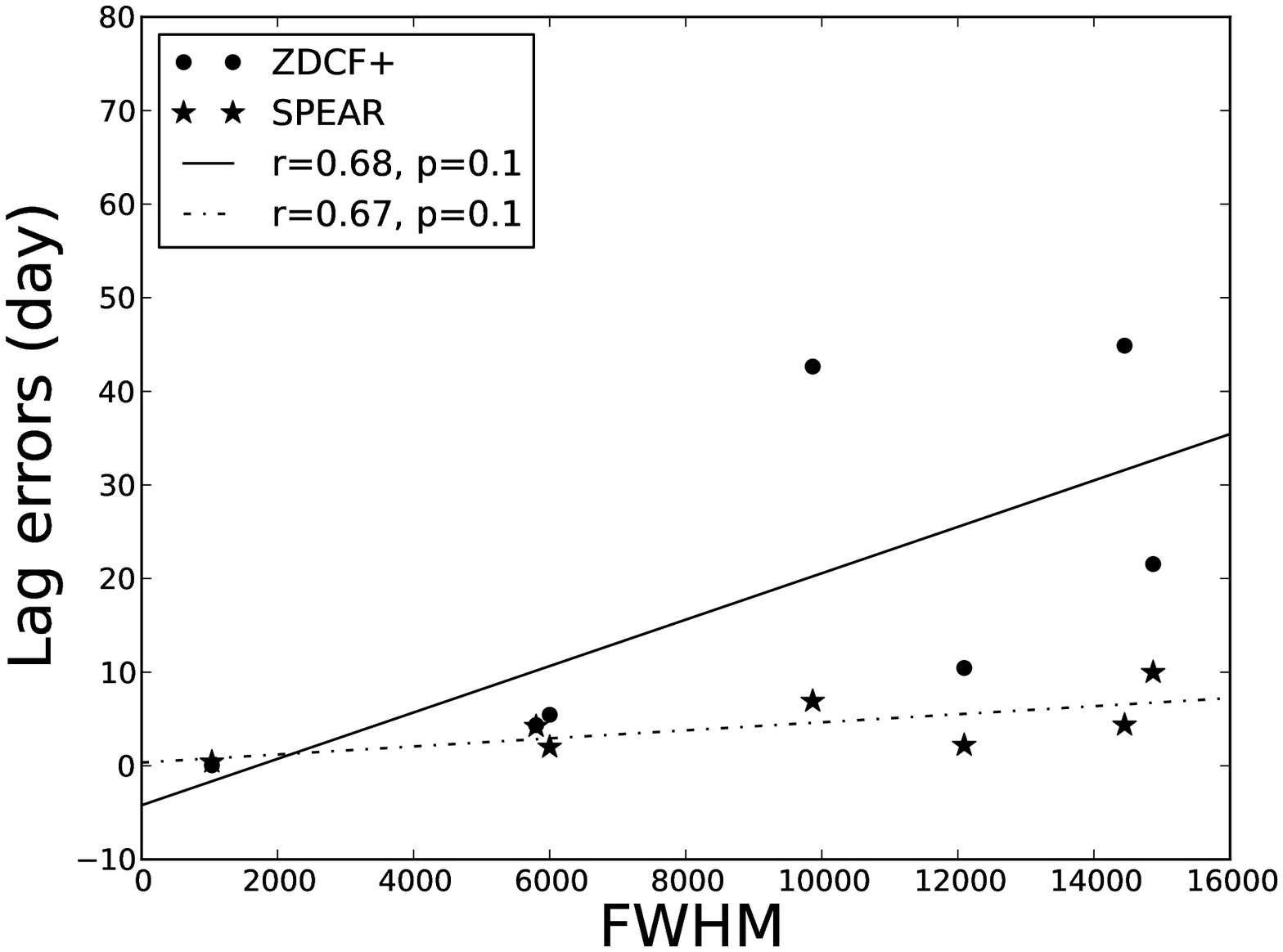}
\caption{ The ratio of the SPEAR and ZDCF time lags as a function of the FWHM of analyzed lines
(upper plot), and the SPEAR and ZDCF lag errors as a function of the FWHM (bottom plot). 
{\bf The horizontal dashed line on upper plot enhances visibility of depicted 
trend}.
The Pearson correlation coefficient $r$ and the p-value are indicated in the upper left corner.} 
\label{corr}
\end{center}
\end{figure} 


  We  also probe the correlation between the sets of calculated time lags,
their errors and emission line widths (see Fig. \ref{corr}). 
For the line widths we use the Full Width Half Maximum (FWHM) of the RMS profiles of H${\alpha}$ and H${\beta}$, 
taken from \citet{Sh13} for Arp 102B, \citet{Sh10} for 3C 390.3, \citet{Sh04} for NGC 5548, and \citet{D09c} for NGC 4051. 
{\bf In the upper panel of Fig. \ref{corr}, depicted 
 trend of higher lag ratio with larger FWHM is not significant 
 (see dashed line with slope 0 at lag ratio 1, passing all the error bars, without increase $\chi^2$ in fitting). }

Obtained  results are just suggesting that deviation between 
two methods is affected by FWHM (see  Fig. \ref{corr},  {\bf bellow panel),  but the trend is not statistically significant.}
 It could be seen that the ZDCF lag errors are larger than SPEAR's.
  Namely, the ZDCF lag errors are calculated  from fiducial distribution and the fiducial interval cannot 
  be narrower than the bin width corresponding to  maximum ZDCF value. Also, the ZDCF method
   bins by equal population and as a result the bins are not equal in time-lag width.

Also, the lag errors are more sensitive to the FWHM in 
ZDCF method than in the SPEAR one (see Fig. \ref{corr},  bottom panel).
 A possible explanation is that the line widths influence strongly the error of time lag measurements. Consequently, this 
increases the uncertainty of the black hole mass.
On the other hand, we found a weak correlation between the FWHM 
and one line SPEAR fitting time lags (without spurious time lag). 
The time lag in cross correlation analysis is the first moment of transfer
function, and it is more sensitive to the autocorrelation function of the
continuum, and because of this we can not find correlations between the ZDCF
time lags and FWHM. {\bf Empirically, it is expected that the scaling coefficient A
in the transfer function is inversely correlated with the ionizing continuum 
flux, where A is defined as  $A=\frac{\left|scale_{1}\cdot scale_{2}\right|}{swid_{1}}$
 \citep{Zu11}. Parameters $scale_{1}$, $scale_{2}$, $swid_{1}$ are 
 scale factors and tophat width obtained from posterior distribution, note that appearance of $swid_{1}$ in denominator
 depends  on SPEAR's input parameters. }

This may cause problems in case of the light curves with the significant long term
trends observed and it could be a reason for correlation between the 
SPEAR time lags (from one line fitting) and FWHM. 
 Also, this should be  tested on large number of simulated light curves, which  will provide more 
insights.



{\bf The time lags from reverberation mapping provide more robust measurement  on the size of the 
BLR comparing with estimating it  from the 5100 \AA\,  luminosity used  in
 empirical virial masses estimate for black holes with H$\beta$ line width  \citep{Kaspi00} 
 and H$\alpha$ line width  \citep{GH05}.} Moreover, our results for the time lags of $\sim$30 ld for Arp 102B and $\sim$80 ld for 3C 390.3
are in good agreement with the estimated disk dimensions for these objects (1000 $R_{\rm G}$ and
1400 $R_{\rm G}$ given by \citet{CH89} and \citet{Flo08}, respectively) if we assume the SMBH masses
of $1.1\times 10^{8} M_{\odot}$ for Arp 102B \citep{Sh13} and $5\times 10^{8} M_{\odot}$ 
for 3C 390.3 \citep{N04,DP12}.

As can be shown by many studies, the choice of the method to calculate time delay should not have to be made apriori, however 
  the unceirtainities depends directly on this decision. In our study we found that the time lags determined by the ZDCF (excluding the large value
of H$\alpha$ of Arp 102B) and SPEAR (excluding large value of H$\alpha$ of 3C 390.3) methods are  consistent with each other as expected. Moreover, the results are consistent 
with previous estimates summarized in Table 4.  {\bf  For example, the consistency of the SPEAR and ICCF 
time lags of 3C 390.3 has been confirmed \citep{Zh13}. These consistencies indicate 
that
our procedures to estimate the time lag by the SPEAR method and by the CCF method are 
reliable.}
 Hereby we note that using different collection of AGNs could bring some other conclusions.
 For example, in the case 
of H$\alpha$ 3C 390.3, that has the worst sampling rate, both methods produced 
largest uncertainties. However, both methods recovered
successfully time lags that are consistent with previous results. 

As for the advantages and disadvantages of the used methods, we can say that the broad
line profiles of Arp 102B, 3C 390.3, and NGC 5548 have produced 'flat-top' ZDCF
curves, while the unfavorable  sampling rate  continuum and emission lines induced 
(Arp 102B, 3C 390.3, and NGC 5548) slow convergence of log-likelihood
functions in the SPEAR method. Also, the broad line profiles could as well
contribute to the slow convergence, as was noticed in case of Mrk 1501 
\citep{Gri12}, which data were well sampled however the
log-likelihood function does not converge. The lines we used for the SPEAR 
two-line fits are pairs of two Balmer lines, which have similar low
ionization levels, and this could affect the fitting results \citep{Zu11}.
Finally, the nature of the non classical BLR of Arp 102B and
3C 390.3 could affect SPEAR and ZDCF to produce larger values for H$\beta$
than for H$\alpha$ time lags. 
 The ZDCF limitation is arising from discrete binning, which implicitly assumes that the spacing 
between the data points is uncorrelated with their observing times. However, if this special sampling
 pattern appears in given light curve, it will be reflected  in spurious fluctuations between consecutive ZDCF points. 
 
  Such fluctuations disappeared only when the ZDCF bin size was enlarged, or when the  data were omitted from the light curves.
   As for limitations of SPEAR method, it
assumes that the continuum variation is a damped random walk and that the line flux should vary in a corresponding way,
 which could be deformed by some other known and unknown mechanisms.

Finally, the characteristics of light curves such as: irregular sampling, correlated errors, and seasonal gaps    
are well-known factors of quality of time-series analysis 
not only in the reverberation mapping campaigns, but also in all the ground-based
time-domain observational studies (including black hole  mass measurements) \citep[see, e.g.][]{Gri08,P11,Zu11}.
 For the black hole mass estimates, among the most crucial factors is accuracy of time lag calculations.
 The expected accuracy of a time delay measurement depends on the number of observation (N)
as follows $\sigma({\tau})\sim \frac{1}{\sqrt{N}}$  
(Horne, 2013, personal communication). Sampling is contributing to the time lag measurements, in the manner as
\citet{ML10} concluded that probing
 the timescales as short as $\frac{\tau}{10}$, ($\tau$  is damping timescale, also called the characteristic timescale)
 and assuming a characteristic redshift of 2 of AGN, the light curves should be sampled every 3 days in the observer’s 
 frame. The common  belief is that the time series used to compute the CCF should
be at least 4 times longer than the lags of interest, and preferably
$\sim10$  times longer, which is well illustrated in \citet{W99}. Note that all our
time series cover periods which are much more than 10 times longer of
calculated time lags. Beside this, the length of time series is also important.
As for an example, more serious problem \citep[see][]{Koen94} than the complexity of the relation between 
light curves could be possible non-stationarity of such relation. 
Non-stationarity could be due to  changes in the lag between the two series. 
This could be analyzed by using very long sets of observations and studying 
segments of the series and comparing the estimate lags. 
Finally, beside frequency domain methods, it is important to use more of cross-spectrum to study
relation between time series \citep[see e.g.][]{Sri09}.

 \section{Conclusion}

Here we present time series analysis of
 the continuum and the H${\alpha}$ and  H${\beta}$ lines  of 
 two well known type 1 AGNs with optical spectra showing double peaked line profiles  (Arp 102B and 
3C 390.3) and two well known broad line AGNs (NGC 5548 and NGC 4051 ). 
We used the ZDCF and the SPEAR method for time lag calculations and
can outline the following conclusions:

(i) {AGNs with broader emission have larger time lag uncertainties in both ZDCF and SPEAR 
methods.
This should be taken into account for the black hole
mass estimates using the virial theoreme. This should be further tested on a larger set of AGNs.

(ii) The ZDCF time lags (without large value of H$\alpha$ of Arp 102B) and SPEAR time lags (without large value
 of H$\alpha$ of 3C 390.3) are  similar.

(iii) Based on (ii) SPEAR and ZDCF produced larger values for H$\beta$ than for H$\alpha$ time lags, 
especially in the case of 3C 390.3, while in the case of NGC 5548 they are closer. 

Also, in the case of Arp 102B  H$\alpha$ line has the time lag which is similar to the time lag of  H$\beta$. 

Finally, we conclude that both SPEAR and ZDCF method give similar results for time lags, that are
consistent with previous measurements  using ICCF methods in the literature. They are  reliable for the time lag analysis of AGN continuum
and emission line light curves, especially in the case of unevenly sampled data. 
 The main advantage of  SPEAR is possibility of rapid increase in the dimensionality of 
the problem: determination lags by fitting  continuum and one line, two lines or even more 
lines (where empirical power spectral distribution of light curves shares the same form with the model)
, while  ZDCF corrects skewness of the sampling distribution of cross correlation 
coefficients by using  z-transform. On the other hand, in the case of SPEAR, 
if the  light curves are intrinsically noisy or the uncertainties are overly underestimated both could affect fitting results, 
while ZDCF method has tendency to overestimate errors.


\section{Acknowledgments}
This work was supported by the Ministry of Education and Science of Republic of Serbia through 
the project Astrophysical Spectroscopy of Extragalactic Objects (176001) and RFBR (grants N12-02-01237a, 12-02-00857a) (Russia) 
and CONACYT research grants 39560, 54480, and 151494 (Mexico). We are grateful 
for the very helpful comments from prof. Keith Horne.
We would like to thank to referees for very useful and detailed  remarks which allowed us to improve our manuscript.

\end{document}